\documentclass{aa}  

\usepackage{graphicx}
\usepackage{txfonts}
\usepackage{caption}
\usepackage{subcaption}
\usepackage{lscape} 
\usepackage{afterpage}
\usepackage{float}
\usepackage{color}

\begin{document}

   \title{Optical Properties of Metal-poor T Dwarf Candidates}

   \author{
  Jerry J.-Y.\ Zhang 
  \inst{1,2}   
     \and  N. Lodieu \inst{1,2}
     \and  E.\ L.\ Mart\'in \inst{1,2}
        }

   \institute{Instituto de Astrof\'isica de Canarias (IAC), 
Calle V\'ia L\'actea s/n, E-38200 La Laguna, Tenerife, Spain \\
       \email{jzhang@iac.es}
       \and
       Departamento de Astrof\'isica, Universidad de La Laguna (ULL), E-38206 La Laguna, Tenerife, Spain
       }

   \date{\today{},\today{}}

 
  \abstract
   {Metal-poor brown dwarfs are poorly understood  
   because they are extremely faint and rare. Only a few candidates 
have been identified as T-type subdwarfs in infrared surveys 
and their optical properties remain unconstrained.}
   {We aim to improve the knowledge of the optical properties 
of T subdwarf candidates to break the degeneracy 
between metallicity and temperature and to investigate 
their atmospheric properties.}
   {Deep $z$-band images of 10 known T subdwarf candidates 
were collected with the 10.4-m Gran Telescopio Canarias. 
Low-resolution optical spectra  for two of them were obtained 
with the same telescope. Photometric measurements of the 
$z$-band flux were performed for all the targets and they 
were combined with infrared photometry in $J$, $H$, $K$, $W1$ 
and $W2$-bands from the literature to  obtain the colours. 
The spectra were compared with solar-metallicity T dwarf 
templates and with laboratory spectra.}
   {We found that the targets segregate into three distinct 
groups in the $W1-W2$ vs. $z-W1$ colour-colour diagram. 
Group I objects are mixed with solar-metallicity T dwarfs. 
Group III objects have $W1-W2$ colours similar to T dwarfs
 but very red $z-W1$ colours. Group II objects 
lie between Group I and III. 
   The two targets for which we obtained spectra are located 
in Group I and their spectroscopic properties resemble 
normal T dwarfs but with water features that are deeper
 and have a shape akin to pure water. 
   }
   {We conclude that the $W1-W2$ vs. $z-W1$ colour-colour 
diagram is excellent to break the  metallicity-temperature 
degeneracy for objects cooler than L-type. 
  A revision of the spectral classification of T subdwarf might be needed in the future, according to the photometric and spectroscopic properties of WISE1810 and WISE0414 in Group III discussed in this work.}

  \maketitle
%
\begin{table*}[htbp]
      \caption[]{Target information ordered by right ascension. 
The coordinates are in degrees under equinox J2000\@. Modified 
Julian dates (MJD) are the epoch dates when the listed positions 
were obtained. If not specified, all the information is from the 
same references as those of the targets. In the rest of the paper,
 we will use WISE\textit{hhmm}, ULAS\textit{hhmm}, or W\textit{hhmm}, 
U\textit{hhmm} as abbreviations.}
         \label{tg}
         
         \begin{tabular}{cccccccc}
            \hline
            \noalign{\smallskip}
           Full name  &  $\alpha$, $\delta$ (deg)
            &  MJD  & $\mu_{\alpha\cos\delta,\ \delta}$ (mas/yr) & Spectral Type \\
            
            \hline

            
    $^{(1)}$WISEA\,J000430.66$-$260402.3
 & $^{(11)}$1.127859$-$26.067652 & $^{(11)}$57234.57 & $^{(11)}$+1$\pm$4, $-$244$\pm$4 & T$2.0$,$^{(11)}$T$4.0$
           \\       
    $^{(2)}$WISEA\,J001354.40+063448.1 & $^{(7)}$3.476628+6.580049& $^{(7)}$55373.51 & +1170$\pm90$, $-$540$\pm90$ & T$8.0\pm0.5$
           \\
    $^{(1)}$WISEA\,J030119.39$-$231921.1 & $^{(11)}$45.331221$-$23.322665 & $^{(11)}$57262.58  & $^{(11)}$+234$\pm$3, $-$119$\pm$4 & T$1.0$
           \\
    $^{(3)}$WISEA\,J042236.95$-$044203.5 &65.656712$-$4.702669\ &58467.54 & +1150$\pm53$, $-$695$\pm55$  & T$8.0\pm1.0$
           \\
    $^{(2)}$WISEA\,J083337.81$+$005213.8 & $^{(12)}$128.408480+0.869044  & $^{(12)}$56658.5 & $^{(12)}$+790$\pm$3, $-$1591$\pm$3 & T$9.0\pm0.5$
           \\
    $^{(4,10)}$ULAS\,J092605.47$+$083516.9 & $^{(8)}$141.522819+8.588039 & $^{(8)}$54120.72 & $-$472$\pm144$, $-$438$\pm146$ & T$4.0\pm1.0$
           \\
    $^{(5)}$ULAS\,J131610.28$+$075553.0 &199.042845+7.931394& $^{(8)}$53885.71 & $-$1012$\pm15$, +103$\pm14$ & T$6.5\pm0.5$
           \\
    $^{(4,10)}$ULAS\,J131943.77+120900.2 &199.932377+12.150072\ & $^{(8)}$54206.05 & $-$525$\pm72$,+111$\pm75$&  T$5.0\pm1.0$
           \\
    $^{(3)}$WISEA\,J155349.96$+$693355.2 & 238.445876+69.567907& 58358.98 & $-$1684$\pm56$, +1348$\pm53$ & $^{(3)}$T$5.0\pm1.0$, $^{(6)}$T4.0
           \\
    $^{(6)}$CWISE\,J221706.28$-$145437.6 &  $^{(9)}$334.276191$-$14.910465 & $^{(9)}$57591.34 &+1637$\pm65$, $-$919$\pm63$ & T$5.5\pm1.2$\\
           
            \hline
         \end{tabular}

    \tablebib{
         (1) \citet{greco2019neowise};
         (2) \citet{pinfield2014subdwarf};
         (3) \citet{meisner2020extremecoldBD}; 
         (4) \citet{murray_burningham2011blueT}; 
         (5) \citet{burningham2014T6.5};
         (6) \citet{meisner2021esdT};
         (7) WISE All-Sky data Release \citet{cutri2012wise};
         (8) UKIDSS LAS DR9 \citet{lawrence2007ulas};
         (9) CatWISE2020 \citet{marocco2021catwise};
         (10) \citet{Burningham2010_47dT};
        (11) \citet{williambest2020_L0-8UKIRTparallax};
        (12) \citet{kirkpatrick2019parallax184TY}.
}

   \end{table*}

\section{Introduction}
   Brown dwarfs (BDs) are substellar objects that have insufficient mass to maintain stable hydrogen thermonuclear burning 
\citep{kumar1963structureBD,burrows1993scienceBD,baraffe1995evolutionBD}. 
Since the discovery of the first BD, Teide 1 in the Pleiades \citep{rebolo1995discovery}, 
and the first BD companion, Gliese\,229\,B \citep{nakajima1995discovery},
thousands of BDs have been identified. 
Characterizing BDs of different ages, temperatures, masses 
and metallicities is critical for tracing substellar evolution 
paths and building ultracool atmospheric models.  BDs are 
extremely faint because of their low surface temperatures and small radii. 
Large-scale deep optical and near-infrared (NIR) sky surveys,
such as 2MASS, the Two Micron All Sky Survey \citep{cutri2003twomasss_point_catalog, skrutskie2006_2MASS}; SDSS, the Sloan Digital Sky Survey \citep{york2000SDSS}; UKIDSS, the United Kingdom Infrared Telescope Infrared Deep 
Sky Survey \citep{lawrence2007ukidss}; WISE, the Wide-field 
Infrared Survey Explorer \citep{wright2010WISE}; and DES, Dark Energy Survey \citep{abbott2019DESdr1,morganson2018DESpipeline,flaugher2015DEScamara},
have been used to search for brown dwarfs.

   Brown dwarfs are classified into different spectral types 
based on their spectral features. After the spectral type M, 
new type sequences L, T, and Y have successively been established 
over the last three decades. The effective temperature 
drops from 2400 K for the earliest L-type to less than 300 K 
for the latest Y-type. \citet{kirkpatrick1999L} and 
\citet{martin1999Lclassification} defined the L-type spectral 
class by noting the strengthening of alkali metal lines, the 
fading of oxide bands and the enhancement of  hydride and water 
bands with respect to M dwarfs. \citet{burgasser2002dT_spec_classification} 
and \citet{geballe2002dT_classification} independently made 
robust T-type classifications based on strong methane bands 
in NIR spectra. Both L- and T-type have been divided into 
into sub-classes from 0.0 (`early type') to 9.5 (`late type'). 
Note that late-M and early-L dwarfs are mixtures of
young massive BDs and old very low-mass dwarf stars 
\citep{dupuy_liu2017dynamical_mass,zhangzenghua2019MLbinary}.  
\citet{delorme2008TY_J0059}, \citet{cushing2011Y}, and \cite{kirkpatrick2011hundredBD_WISE} found 
the first examples of Y dwarfs and proposed that ammonia 
absorption in the $H$-band could be a characteristic of this 
class, which is cooler than T. So far, only tens of Y dwarfs have 
been discovered and this class extends only to the Y2 subclass. 
  
   Metal-poor dwarfs are also called subdwarfs. 
  Their chemical composition remains pristine except for 
the fusion of some light elements such as deuterium 
\citep{chabrier2000deuterium} and lithium for the most 
massive ones close to the substellar limit 
\citep{rebolo1992spectroscopy_lithum,basri1996lithium}. 
   They have been continuously fading and cooling down 
for billions of years with increasingly degenerate cores 
and no significant thermonuclear reactions. 
 Thousands of M subdwarfs  
\citep{zhangshuo2019sdM_identification_sample} and 
dozens of L subdwarfs \citep{zhangzenghua2018subL} have 
been identified, but only a handful of T subdwarf 
candidates have been announced. 

The optical properties of T subdwarf candidates have 
not been explored yet and may be crucial for 
breaking the metallicity and temperature degeneracy, as 
has been done for M and L subdwarfs 
\citep{zhang2017six_sdL_classification,lodieu2019sdM}. 

In this paper, we present observations of 10 metal-poor 
T subdwarf candidates that allow us to obtain photometry 
in the $z$-band for all of them and optical  spectroscopy 
for two of them. Section~\ref{observation} describes the 
observation protocols and data reduction processes. 
Sections~\ref{astsec}, \ref{photsec} and \ref{specsec} show 
the results of astrometry, photometry, and spectroscopy, as 
well as our analysis. Section~\ref{FHTrend} describes a new potential metallicity indicator the $z-W1$ colour, and discusses possible changes required to the current T subdwarf classification. The last section~\ref{conclusion} provides a summary and discussion 
of the impact of this research and future applications. 

         \begin{table*}[htbp]
      \centering
      \caption[]{Logs of photometric and spectroscopic 
observations. The MJD indicates the time at the middle of the observation.}
         \label{Obs}
         
         \begin{tabular}{cccccc}
            \hline
            \noalign{\smallskip}
            Name  &         Observation MJD &  Instrument   & Total exp. (s)    & Std. \ star  \\
            
            \hline

  W0004 & 59810.14, 59811.21 &OSIRIS Long Slit R500R& 7200 &Ross 640  
           \\
 W0301& 59844.13, 59871.11 &OSIRIS Long Slit R500R & 7200 & Feige 110
 \\
            
            \hline
    W0004 & 59811.16 &OSIRIS $z$-band acquisition & 90 &   \\
 
    W0013 &59810.19 &OSIRIS $z$-band & 2500 & 
    \\

     W0301 & 59844.13 &OSIRIS $z$-band acquisition & 90 &  \\
           
    W0422 &59844.18 &OSIRIS $z$-band & 2000 & 
           \\
    W0833 & 59971.50 &OSIRIS$+$ $z$-band & 2100 & 
           \\
    U0926 & 59971.54 &OSIRIS$+$ $z$-band & 2350 & 
           \\
    U1316 &59725.91\ &OSIRIS $z$-band & 2150 & 
           \\
    U1319 &59725.94\ &OSIRIS $z$-band & 2000 & 
           \\
    W1553 &59725.02\ &OSIRIS $z$-band & 1700 & 
           \\
    W2217 &59826.03 &OSIRIS $z$-band & 2000 & 
           \\
           
            \hline
         \end{tabular}
   \end{table*}


   \section{Observations}
\label{observation}

Our sample consists of 10 T subdwarf candidates from 
the literature that are observable from the  
Roque de los Muchachos Observatory on the island of La Palma 
(Spain). \citet{Burningham2010_47dT,burningham2014T6.5} 
discovered ULAS0926, ULAS1316 and ULAS1319 from the Large 
Area Survey (LAS) of UKIDSS based on their blue NIR colours. 
Their ultracool subdwarf status was confirmed by NIR
spectroscopy.  \citet{pinfield2014subdwarf} discovered the 
two late-T subdwarf candidates WISE0013 and WISE0833 from 
$W2$-only detection in WISE and classified them based on 
their halo kinematics, metal-poor NIR colours and NIR 
spectroscopy. \cite{greco2019neowise} identified WISE0004 
and WISE0301 from the NEOWISE proper motion survey and 
classified them as T subdwarfs after NIR spectroscopic 
follow-up. \citet{meisner2020extremecoldBD,meisner2021esdT} 
identified WISE0422, WISE1553 and WISE2217 from the 
Backyard Worlds: Planet 9 Citizen Science Project, which 
visually selects the candidates with high 
proper motions. Target information is listed in Table~\ref{tg}.

\subsection{Observation details}
\label{obsdetail}

We collected $z$-band photometry of these candidates using 
the imaging mode of Optical System for Imaging and 
low-Intermediate-Resolution Integrated Spectroscopy (OSIRIS 
\citealt{cepa2000osiris}) on the 10.4 m Gran Telescopio 
Canarias (GTC) on La Palma. We requested 
four-point dithering pattern with ten 50 s exposures at each 
point. Our weather constraints were seeing less than 0.9",
 grey nights and clear sky. In practice, the support 
astronomer adapted the numbers of exposures for some of the 
objects to the observing conditions, resulting in 
slightly different total integration times, see Table~\ref{observation}. 

We also made spectroscopic observations using the OSIRIS long 
slit mode of the two brightest targets, WISE0004 and WISE0301, 
under the same weather constraints as for the imaging mode.  
Before the spectroscopic exposures we took three 30 s 
acquisition images with $z$-band filter and made use of them 
for the $z$-band photometry.  We used the R500R grism and 1.0" slit, resulting in a resolving power $R \approx 350$. We 
selected parallactic angle to prevent flux lost due to 
atmospheric refraction.  We created two observing blocks for 
two nights, each with two spectra of 1800 s shifted along the 
slit by 10$^{\prime\prime}$, and for each observing block we observed a 
spectroscopic standard star using the same grism and slit width 
with and without the $z$-band filter. 
\citet{zapatero2018Pleiades_leastmassive} first introduced this 
method of using the $z$-filtered standard spectrum to eliminate 
the second-order contamination (approximately from $9600$ \AA\ 
to $9800$ \AA) from the blue light (from $4800$ \AA\ to 
$4900$ \AA), which is a problem for all the OSIRIS red grisms. 
Our targets have negligible flux in the blue part 
so there is no second-order contamination.

 We note that ULAS0926 and WISE0833 were observed after the 
CCD  replacement of OSIRIS on 2022 December 12. The main 
differences is that the new camera, called OSIRIS$+$, has  
one monolithic CCD instead of two, and that it improves the 
sensitivity specially in the blue wavelength range ($<5500$ \AA).  
A summary of all observations is provided in Table~\ref{observation}.

\subsection{Data reduction}
\label{datareduction}

We reduced the OSIRIS $z$-band imaging data using IRAF 
\citep{tody1986iraf}.  First, the sky frame of each position 
in the four-point dithering pattern was created by median-scaling 
and median-combining  all the images at the other three positions, 
but two thirds of the highest pixels were rejected, as we regarded 
the lowest to represent the optimal sky value. The corresponding 
sky frame was then subtracted from all the images with respect to each 
position. Finally all the sky-subtracted images were aligned 
and average-combined, but two highest and two lowest pixels were rejected.

We reduced the OSIRIS spectroscopic data with \textit{PypeIt} 
\citep{pypeit:zenodo,pypeit:joss_pub}.  \textit{PypeIt} can 
generate calibration frames, subtract the sky, 
extract the object spectra, do the wavelength calibration, generate 
the sensitivity function and use it to do the flux calibration, 
coadd the spectra and finally perform the telluric correction. 
We tuned  \textit{PypeIt} to find the object in the y-pixel region 
between 1200 to 1600, because our objects are red and have very 
little flux below 8000 \AA .  

The sensitivity function for instrumental response correction
was generated from the standard star spectrum. We spliced 
the part before $9000$ \AA\ of non-$z$-filtered standard spectrum 
and the part after $9000$ \AA\ of the $z$-filtered standard spectrum 
to obtain the standard spectrum covering the whole wavelength 
range without second-order contamination; the method is 
explained in Section~\ref{obsdetail}.  

We fit a low-order polynomial to the observed spectrum.  
The $9150$ \AA\ to $9900$ \AA\ water absorption wavelength region
 of the spectra was masked in the fit. Then the pipeline used 
the Maunakea telluric grids to correct the spectra from the 
telluric contribution.

\section{Astrometry}
\label{astsec}

\subsection{Object recognition}
  We created the world coordinate system (WCS) for the OSIRIS 
images using \textit{Astrometry.net} \citep{dustin2010astrometry.net}, 
which extracts stars and solves the WCS by matching subsets of 
four stars to the pre-computed 4200 series index based on the 
2MASS catalogue. We requested \textit{Astrometry.net} to 
match only the field around the telescope's pointing position within 
one degree. 
  
  2000-second exposure of GTC ORISIS $z$-band goes deep down 
to 24.5 mag (3-$\sigma$, 1.0" seeing and dark condition), 
such that there are no counterpart $z$-band images with a similar 
depth requiring us to exclude the contamination and identify the objects. 
Therefore, to avoid false recognition we projected the expected 
position region of the objects in the OSIRIS images based on their 
proper motions and their associated errors, the positions and the 
epoch of the corresponding position, and the time differences 
between the previous epoch and the epoch our OSIRIS observation.  
Table~\ref{tg} lists the proper motions from the literature.  
We drew a circle with a radius equal to the error  centred on 
the projected position for each target.

We detected beyond doubt WISE0004, WISE0013, WISE0301, WISE0422, 
WISE0833, and ULAS1316 at their projected positions.  However, there 
are no detections at the projected positions of ULAS0926, ULAS1319 
and WISE2217. For ULAS0926 and ULAS1319, the proper motions given
 by \citet{Burningham2010_47dT} are over-estimated because the 
epoch differences are too short; and for WISE2217, the projection 
was slightly displaced from the object. The objects are indicated by the 
red arrows, see the bottom three cutouts in Figure~\ref{fig:three graphs}.

\subsection{Proper motion revision}
For the latter three objects, we used the pixel positions with the centroid errors given by the IRAF task \textit{imcentroid} and the WCS from 
\textit{Astrometry.net} to get their coordinates in our observations. To be sure about 
the astrometric accuracy, we calculated the rms of the fit of 
the coordinate transformation using the index pixel positions 
and field pixel positions in X and Y in corr.fits files generated by \textit{Astrometry.net}. The rms is the standard deviation of the mean of the pixel deviation of the matched reference stars (match$\_$weight $>0.99$). For the three objects the rms are at sub-pixel level.    We derived proper motions using two epochs: the first epoch in Table~\ref{tg} and the second epoch of our observations. \citet{murray_burningham2011blueT} gave the first epoch position errors of ULAS0926 and ULAS1319; and CatWISE provided those for WISE2217.  The results are given in Table~3.

\begin{table}[H]
\centering
      \caption[]{Revised astrometry of the three objects with the astrometry rms, the centroid errors, and the pixel sizes.}
         \label{astrometry}
         ULAS0926 at MJD 59971.54
         \\
         \vspace{1mm}
         \begin{tabular}{cc}
            \hline
            $\alpha$, $\delta$ (deg)& 141.522478+8.587854
            \\
            rms$_{X, Y}$ (pix) & 0.313, 0.187
            \\
            err$_{X, Y}$ (pix) & 0.061, 0.063
            \\
            scale (mas/pix) & 254.92  (OSIRIS+)
            \\
            $\mu_{\alpha\cos\delta,\ \delta}$ (mas/yr) &$-76\pm9$, $-42\pm8$
            \\
            \hline
         \end{tabular}
\vspace{2mm}

         ULAS1319 at MJD 59725.91
         \\
         \vspace{1mm}
         \begin{tabular}{cc}
            \hline
            $\alpha$, $\delta$ (deg)& 199.932017+12.150141
            \\
            rms$_{X, Y}$ (pix) & 0.619, 0.332
            \\
            err$_{X, Y}$ (pix) & 0.053, 0.065
            \\
            scale (mas/pix) & 258.30
            \\
             $\mu_{\alpha\cos\delta,\ \delta}$ (mas/yr)&  $-85\pm13$, $16\pm9$
            \\
            \hline
         \end{tabular}
\vspace{2mm}

         WISE2217 at MJD 59826.03
         \\
         \vspace{1mm}
         \begin{tabular}{cc}
            \hline
            $\alpha$, $\delta$ (deg)& 334.278402$-$14.912140
            \\
            rms$_{X, Y}$ (pix) & 0.375, 0.396
            \\
            err$_{X, Y}$ (pix) & 0.126, 0.152
            \\
            scale (mas/pix) & 259.25
            \\
            $\mu_{\alpha\cos\delta,\ \delta}$  (mas/yr)& $1257\pm44$, $-986\pm44$
            \\
            \hline
         \end{tabular}
   \end{table}

\section{Photometry}
\label{photsec}

We performed aperture photometry on the reduced images of 
those objects using the \textit{Astropy} package 
\textit{photutils}. The aperture radius is fixed to be 
1", and the background residual after the sky subtraction 
is from the median value in an annulus of 3.5" inner 
radius and 5.5" outer radius.  We did the same photometric 
measurements on an extra set of  3--6 neary stars of  
Pan-STARRS magnitudes  between 18 mag and 20.3 mag  
(AB system, \citet{oke1974ABsystem,chanbers2016panstarrs}). 
We used their $z$-band magnitudes to find the zero-points. 
We assumed there to be Poisson errors and the uncertainties 
from zero-point fit. The Poissonian error of aperture count $A$ 
from the average combination of $N$ exposures with same 
exposure time is $\sqrt{A/N}$. The final 
error is simply the square root of the quadratic sum of the 
Poissonian error and the zero-point uncertainty.

\subsection{Two extra objects: WISE0414 and WISE1810}

For reference sources, we added two extra extreme T subdwarf 
candidates WISEA\,J041451.67$-$585456.7 and WISEA\,J181006.18$-$101000.5, 
which were reported and spectroscopically characterized 
in the NIR by \citet{Schneider2020W0414_W1810}. They have 
metallicities [Fe/H] $\approx-1$ dex and $\leq-1$ dex, 
respectively. \citet{lodieu2022W1810} collected GTC OSIRIS 
$z$-band photometry and spectroscopy allowing them to place 
improved constraints on the metallicity of WISE1810 ($-1.5\pm0.5$ dex).  
We also found that WISE0414 has $z$-band detection in the Dark 
Energy Survey (DES),  see Figure~\ref{W0414_illu}. The DES also 
recorded its magnitudes in  the $r$-band at 26.54$\pm$1.39 mag; and in 
the $i$-band at 25.71$\pm$1.13 mag.

\subsection{Correction for Pan-STARRS $z_{AB}$ magnitude}

The $z$ filter of Pan-STARRS\footnote{Transmission profile 
of the PS1 $z$ filter: 
\url{http://svo2.cab.inta-csic.es/theory/fps/index.php?id=PAN-STARRS/PS1.z&&mode=browse&gname=PAN-STARRS&gname2=PS1#filter}.} 
and the $z'$ filter of SDSS used by OSIRIS\footnote{Transmission 
profile of the Sloan $z'$ filter on OSIRIS: 
\url{http://svo2.cab.inta-csic.es/theory/fps/index.php?id=GTC/OSIRIS.sdss_z&&mode=browse&gname=GTC&gname2=OSIRIS#filter}}  
are very different; the latter in particular allows the 
flux beyond $9300\AA$ pass through and the former does 
not. Moreover, the SED profiles of T dwarfs are quite 
inclined around this wavelength. Thus we are obliged to apply 
an offset to the GTC OSIRIS magnitudes to get the magnitudes 
under the Pan-STARRS AB system.

 We synthesized photometry and computed the offsets between
 OSIRIS magnitude $z'_{AB}$ and Pan-STARRS magnitude $z_{AB}$ 
against the NIR spectral types of dwarfs based on two filter
 profiles and L dwarf optical templates \citep{kirkpatrick1999L}; 
the T dwarf optical templates  
\citep{burgasser2002dT_spec_classification,burgasser2003dT_optical}, 
and Y dwarf theoretical models \citep{morley2014Ywaterclouds}.
 The Y dwarf models have gravity $\log g = 4.5$, sedimentation 
efficiency $F_{\rm sed}=5$, cloud cover $h=50\%$, and effective 
temperature $T_{\rm eff}$ from 450K to 200K. We converted the 
$T_{\rm eff}$ of the Y dwarf to spectral types 
\citep{Schneider2015HubbleWISE_BD, cushing2021W1828}. 
Not all the dwarf templates are metal-poor, but we consider
 that, within a $1000\AA$ bandwidth in $z$-band, the difference 
in the  SED slopes are so small that we can also apply this to 
our metal-poor samples.  Because WISE0414 was detected in DES 
we repeated the procedure for the DES 
filters.\footnote{Transmission profile of DECam $z$ filter 
of DES: \url{https://noirlab.edu/science/node/41112}.}  
The DES $z$ filter has more transmitivity than the SDSS $z'$ 
filter in the red part.  Table~\ref{PS1_SDSS} lists the results.

We did a consistency check on the calculated offset 
values. The two brightest objects have detections in 
Pan-STARRS DR1 (PS1): WISE0004 which is recently classified 
as a T2 or T4 and WISE0301 which is classified as a T1 
\citep{williambest2020_L0-8UKIRTparallax}. Both are 
early-T type dwarfs. The differences between Pan-STARRS 
AB magnitudes and the GTC magnitudes $z_{AB}-z'_{AB}$ are 
$0.58\pm0.04$ mag and $0.47\pm0.05$ mag, respectively. We 
recalculated the offsets using their ORISIS spectra 
(presented in Section~\ref{specsec}) and found $0.61$ mag 
and $0.52$ mag, respectively.  If we assume an uncertainty 
of $\pm1.0$ in the sub-class in the spectral type of these 
two objects, both results are consistent with Table~4, 
which are from $0.68$ to $0.87$ mag and from $0.43$ to 
$0.72$ mag, respectively. We computed the offset of 
WISE1810 using its OSIRIS spectra (obtained by \citet{lodieu2022W1810} 
and presented in Section~\ref{specsec}). WISE1810 has 
$z_{AB}-z'_{AB}=0.54$ mag, which is consistent in Table~\ref{PS1_SDSS}
 with its spectral type estimation ($0.41$ to $0.68$ mag; 
T0.0$\pm$1.0, \cite{Schneider2020W0414_W1810}). We also 
computed the offset for the Y dwarf WISE\,J173835.53+273259.0
 with OSIRIS spectrum obtained by \citet{martin2023Yoptical}.
 It has an estimated effective temperature of 300--450 K 
\citep{kirkpatrick2011hundredBD_WISE}, which corresponds to 
Y0--Y2, although \citet{kirkpatrick2011hundredBD_WISE} 
classified it as a Y0 in the NIR. The offset turns out to be 
$0.17$ mag, which is in line with the predicted results 
in Table~\ref{PS1_SDSS} ($-0.12$ to $0.59$ mag).  The mid-T 
types reach the maximum offset around 0.9 mag for both the GTC 
and the DES.
 
 \begin{table}[H]
\centering
      \caption[]{$z$-band magnitude offsets between GTC 
OSIRIS $z'_{AB}$ and Pan-STARRS1  $z_{AB}$;  DES $z_{DES}$ and 
$z_{AB}$ of different spectral types from late-L to Y from 
synthesized photometry. }
         \label{PS1_SDSS}
         \begin{tabular}{ccc}
            \hline
            Spec. type &  $z_{AB}-z'_{AB}$  &  $z_{AB}-z_{DES}$
            \\
            \hline
            
            L6.0 & $0.31$ & $0.40$
            \\
            L7.0 & $0.43$ & $0.51$
            \\
            L8.0 & $0.41$ & $0.50$
            
            \vspace{0.8mm}
            \\
            T0.0 & $0.43$ & $0.52$
            \\
            T1.0 & $0.68$ & $0.72$
            \\
            T2.0 & $0.72$ & $0.81$
            \\
            T3.0 & $0.72$ & $0.79$
            \\
            T4.5 & $0.87$ & $0.92$
            \\
            T5.0 & $0.86$ & $0.91$
            \\
            T6.0 & $0.86$ & $0.89$
            \\
            T7.0 & $0.81$ & $0.80$
            \\
            T8.0 & $0.82$ & $0.83$
            
            \vspace{0.8mm}
            \\
            Y0.0 & $0.59$ & $0.52$
            \\            
            Y1.0 & $0.38$ & $0.35$
            \\
            Y2.0 & $-0.12$ & $0.06$
            \\
            \hline
         \end{tabular}
\end{table}

For the two brightest objects, WISE0004 and WISE0301, we 
simply adopted their Pan-STARRS AB magnitudes. For WISE1810, 
we use the correction calculated using its spectrum. For 
the rest, we applied the offsets in Table~\ref{PS1_SDSS} to 
all the objects according to their estimated NIR spectral 
types in Table~\ref{tg}.  The uncertainty of the correction 
comes from both the uncertainty of the spectral type and the 
coarseness of the spectral-type step in Table~\ref{PS1_SDSS}. 
The final $z_{AB}$ error is the square root of the quadratic 
sum of this uncertainty and the photometric error of $z'_{AB}$. 
The photometry results are shown in Table~\ref{z}.

\begin{table*}[htbp]
    \centering
        \caption[]{GTC $z$-band magnitudes $z'_{AB}$ of all the targets, except WISE0414 is from DES. Pan-STARRS AB magnitudes $z_{AB}$ were calculated according to Table~\ref{PS1_SDSS}, except for WISE0004 and WISE0301 for which we adopted the magnitudes from the PS1 catalogue.  Published NIR photometry (from the same references shown in Table~\ref{tg} unless specified) are also listed.  ULAS1316 was moving onto a galaxy, so we adopted the $Y$, $J$ photometry in epoch 2006 rather than in 2010, and we questioned its $W1$ and $W2$ photometry.}
         \label{z}
    \resizebox{\textwidth}{!}{\begin{tabular}{ccccccccc}  
    \hline
        Name & $J$ & $W1$ & $W2$ & $Y$ & $H$ & $K$ & $z'_{AB}$ & $z_{AB}$ \\ \hline
        W0004 & 16.18$\pm$0.02 & 15.21$\pm$0.04 & 14.13$\pm$0.04 & ~ & 15.59$\pm$0.13 & > 15.52 & 19.53$\pm$0.02 &  20.12$\pm$0.03 \\ 
        W0013 & 19.75$\pm$0.05 & > 18.31 & 15.04$\pm$0.11 & ~ & 20.02$\pm$0.05 & ~ & 23.39$\pm$0.05 &  24.21$\pm$0.05 \\
        W0301 & 16.63$\pm$0.02 & 14.83$\pm$0.03 & 14.04$\pm$0.04 & ~ & 15.80$\pm$0.16 & 15.58$\pm$0.23 & 19.68$\pm$0.02 &  20.14$\pm$0.05 \\ 
        $^{(1)}$W0414 & 19.63$\pm$0.11 & 16.71$\pm$0.03 & 15.29$\pm$0.03 & 21.62$\pm$0.17 & & > 18.82&  &23.35$\pm$0.23\\
        W0422 & 19.43$\pm$0.23 & 18.82$\pm$0.41 & 16.02$\pm$0.08 & > 19.97 & ~ & $K_s$ > 18.89 & 23.94$\pm$0.08 & 24.64$\pm$0.14\\ 
        W0833 & 20.28$\pm$0.10 & > 18.35 & 14.96$\pm$0.10 & 20.43$\pm$0.21 & 20.63$\pm$0.10 & $^{(4)}$ > 18.20 & 23.57$\pm$0.09  &  24.27$\pm$0.15 \\
        U0926 & 18.57$\pm$0.02 & $^{(3)}$18.06$\pm$0.13 & $^{(3)}$17.10$\pm$0.18 & 19.86$\pm$0.15 & 18.69$\pm$0.01 & 19.14$\pm$0.25 & 22.29$\pm$0.03 &  23.08$\pm$0.07 \\
        U1316 & 19.29$\pm$0.12 & $^{(3)}$16.54$\pm$0.04 & $^{(3)}$15.95$\pm$0.06 & 20.00$\pm$0.14 &  $^{(4)}$ > 18.80 & $^{(4)}$ > 18.20 & 22.88$\pm$0.04 &  23.72$\pm$0.05  \\
        U1319 & 18.90$\pm$0.05 & ~ & ~ & 20.39$\pm$0.05 & 18.90$\pm$0.15 & 19.41$\pm$0.10 & 22.78$\pm$0.04 &  23.64$\pm$0.04 \\ 
        W1553 & $^{(5)}$19.17$\pm$0.03 & 17.07$\pm$0.03 & 15.62$\pm$0.03 & ~ & $^{(5)}$18.87$\pm$0.05 & $^{(5)}$19.24$\pm$0.03 & 22.08$\pm$0.02 &  22.94$\pm$0.02  \\ 
        $^{(1,2)}$W1810 & 17.29$\pm$0.04 & 13.92$\pm$0.03 & 12.58$\pm$ 0.04 & 18.94$\pm$0.13 & 16.52$\pm$0.03 & 17.10$\pm$0.17 & 20.15$\pm$0.08 & 20.69$\pm$0.08 \\
        W2217 & 20.66$\pm$0.02 & 17.43$\pm$0.08 & 15.78$\pm$0.06 & ~ & $^{(5)}$20.66$\pm$0.06 & $K_s$ > 18.20 & 23.52$\pm$0.08 &  24.36$\pm$0.09  \\ 
        \hline
    \end{tabular}}
     \tablebib{(1): \cite{Schneider2020W0414_W1810};  
(2): \cite{lodieu2022W1810}; (3): CatWISE \citep{marocco2021catwise}; 
(4): ULAS \citep{lawrence2007ulas}; (5): \citet{meisner2023coldoldBD}.}
\end{table*}

\begin{figure*}[htbp]
    \centering
    \includegraphics[width=\textwidth]{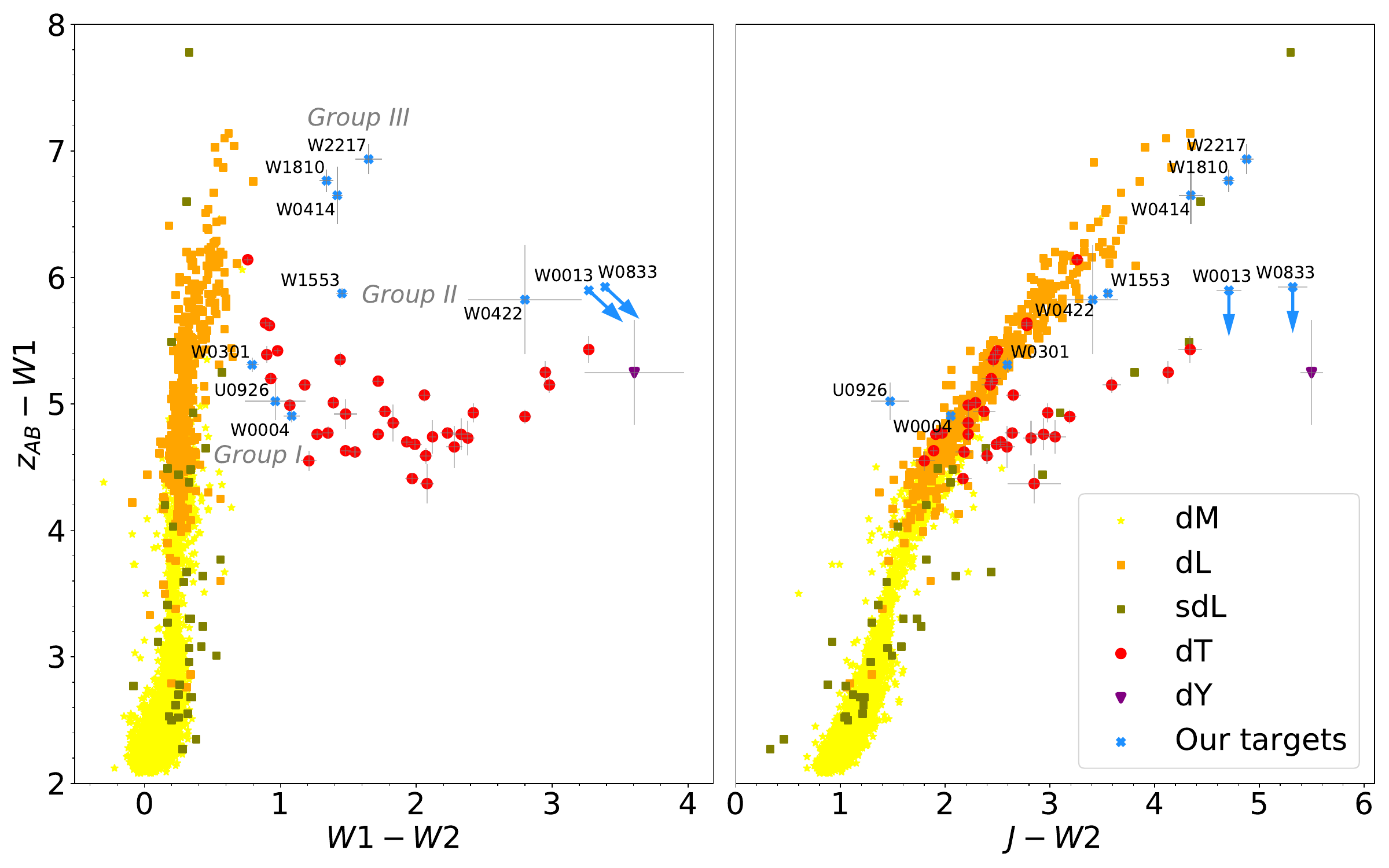}
    \caption{$W1-W2$ vs. $z_{AB}-W1$ and $J-W2$ vs.
 $z_{AB}-W1$  colour--colour diagrams of all the T subdwarf 
candidates in Table~\ref{z} except  ULAS1319, which has 
neither $W1$ nor $W2$ magnitudes, and ULAS1316 whose 
photometry was contaminated by a background galaxy. For
 WISE0833 and WISE0013, we used arrows to indicate the 
lower limit of $W1$ magnitude. We also included normal M, 
L, T sequences, sub-L dwarfs and a Y dwarf. All the T dwarfs, the
Y dwarf and our objects have error bars.  In the first 
diagram the candidates clearly separate into three groups.}
    \label{colourcolour}
\end{figure*}


\begin{figure*}[htbp]
\centering
    \includegraphics[width=0.95\textwidth]{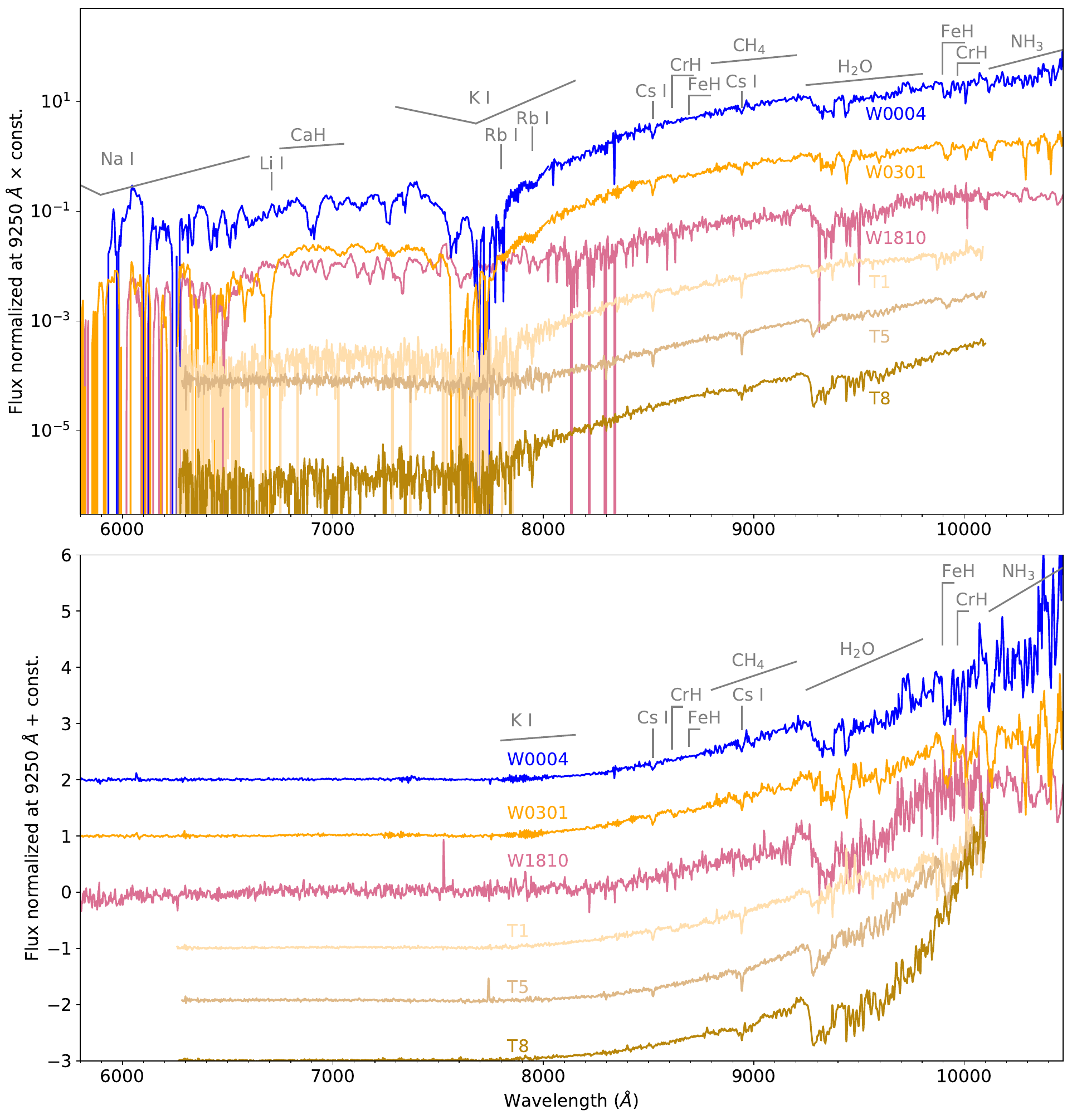}
    \caption{Full optical spectra from 5800 $\AA$ to 
10470 $\AA$ normalized at 9250 $\AA$ in  a logarithmic and 
a linear scale of the two  T subdwarf candidates and WISE1810, 
with the alkali atomic lines (vertical lines or pressure
 broadened V-shaped lines), molecular bands (horizontal 
lines) and band heads (vertical lines with a dash). In the logarithmic scale plot, we smoothed the parts below 8000 $\AA$ of these three spectra to not let the noise block our sight.}  We 
plotted optical spectra of the T1 standard SDSS0837, the 
T5 standard 2MASS0755 and the T8 standard 2MASS0415 
\citep{burgasser2003dT_optical} for comparison. 
    \label{optspec}
\end{figure*}

\subsection{$z$-band magnitudes and colour--colour diagrams}
\label{secccd}
    Table~5 shows the GTC $z$-band magnitudes  $z'_{AB}$ and 
the offset-corrected Pan-STARRS AB magnitude $z_{AB}$ as 
well as the $Y, J, H, K, W1, W2$ magnitudes or magnitude limits for all the targets in our sample. $z_{AB}$ of WISE0414 is corrected from its DES $z$ magnitude, and 
its estimated spectral type T0.0$\pm$1.0, the same as WISE1810 
\citep{Schneider2020W0414_W1810}. We noticed that our 
GTC $z'_{AB}$ magnitude of WISE1553 is in agreement with  
$22.17\pm0.21$ mag \citep{nidever2018_survey_w1553zmag} 
from Mosaic3 with a $z$ filter similar to that of DES of
 Kitt Peak Nicholas Mayall Telescope
    \citep{Dey2016kittpeak_mosaic3}. We took the offset 
between our $z'_{AB}$ and DES $z$ into account (shown in 
Table~\ref{PS1_SDSS}). Our $z'_{AB}$ magnitude for ULAS0926 is
 consistent at the 1-$\sigma$ level with the $z'_{AB}$ magnitude 
$22.16\pm0.09$ mag converted by \citet{Burningham2010_47dT} 
from the $z$ magnitude of ESO Multi-Mode Instrument (EMMI). 
 For ULAS1316,  we adopted the $Y, J, H, K$ photometric 
measurement at epoch  2006.41 rather than those
 at epoch 2010.24, because ULAS1316 gradually
 moved onto a background galaxy during that time and thus the photometry had more contamination. For the same reason, the $W1$ and $W2$ magnitudes are questionable.  For 
ULAS1319, there is no $W1$ nor $W2$ magnitude because the WISE satellite was not able to resolve it from a 
nearby bright star WISE\,J131943.68+120907.0 ($W1=12.74$ mag).

Figure~\ref{colourcolour} shows the colour--colour diagrams of $W1-W2$ vs.\ 
$z_{AB}-W1$ and $J-W2$ vs.\ $z-W1$ for all the objects except 
ULAS1316 and ULAS1319, with extra 8508 M dwarfs, 
800 L dwarfs and 42 T dwarfs from the Pan-STARRS1 3$\pi$ Survey 
\citep{best2018PS1_3pi}; one Y0 dwarf WISEP\, J173835.52+273258.9 
discovered by \citet{cushing2011dY_WISE} with both $W1$ and 
$W2$ detection and was detected in $z$ using GTC OSIRIS  
\citep{lodieu2013Yoptical}; 39 L  subdwarfs with photometric 
errors less than 0.2 mag in $z, J, W1$, and $W2$-bands   
\citep{zhangzenghua2018subL}.

\begin{figure*}[htbp]
\centering
    \includegraphics[width=\textwidth]{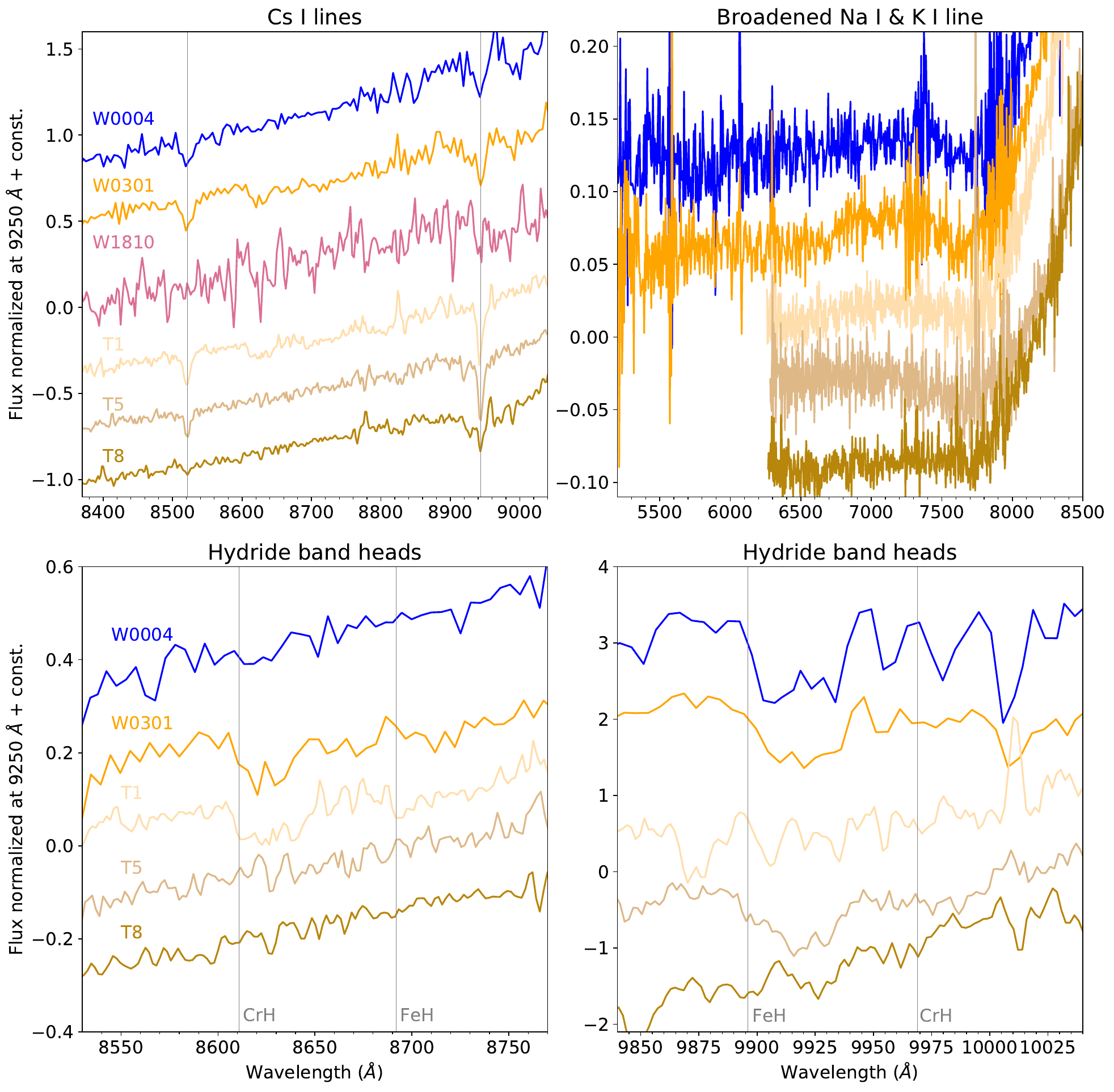}
    \caption{Optical spectra normalized at 9250 $\AA$ 
on a linear scale in the alkali metal line region (Cs, Na and 
K, two top panels), and the hydride band regions (FeH and CrH, 
two bottom panels) of the two T subdwarf candidates and
 WISE1810 compared with three T dwarf templates. 
We did not plot the spectrum of WISE1810 because of its
low signal-to-noise ratio.}
    \label{component}
\end{figure*}

\citet{kirkpatrick2011hundredBD_WISE} showed observationally that the
$W1-W2$ colour is a very powerful aid to distinguishing solar-metallicity T and Y dwarfs from M and L dwarfs; the colour becomes redder 
monotonically towards later types, and the slope of 
colour vs.\ spectral type relation curve increases dramatically 
after passing the L/T transition point. The $J-W2$ colour, however, 
has some degeneracy among late-L and early-T dwarfs.

In the first $W1-W2$ vs.\ $z_{AB}-W1$ colour--colour diagram,
 all of our objects have $W1-W2\geq0.79$ and split into 
three quite distinctive groups in 
$z_{AB}-W1$ colours: Group I is mixed with the normal T dwarf 
sequence, Group III has an extremely red $z_{AB}-W1$ colour 
$> 6.5$ mag and Group II lies in between Group I and Group III.  
The separation between the three groups in the second $J-W2$ vs.\ 
$z_{AB}-W1$ colour--colour diagram is not as clear as that in 
the first diagram. We explain in 
Section~\ref{FHTrend} that $z-W1$ can be a good metallicity 
indicator for objects with $W1-W2$ colours similar to those of T dwarfs.

\section{Spectroscopy}
\label{specsec}
Our two brightest targets, WISE0004 and WISE0301, lie in 
Group I at the beginning of the T dwarf sequence in the $W1-W2$ 
vs.\ $z_{AB}-W1$ colour--colour diagram. In this section we 
present and discuss their optical spectra (shown in Figure~\ref{optspec}), 
\ref{component}, and \ref{lab}.   To compare the T subdwarf 
with normal T dwarf, we used three spectra of T standards 
from the Low Resolution Imaging Spectrograph (LRIS) on Keck I, the 
T1 standard SDSS\,J083717.31$-$000018.0, the T5 standard 
2MASS\,J07554795$+$2212169 and the T8 standard 2MASS\,J04151954$-$0935066 
\citep{burgasser2003dT_optical}. We also included the GTC 
OSIRIS spectrum of WISE1810, which is from Group III 
\citep{lodieu2022W1810} in the comparison,  but we shall discuss 
it in detail in Section~\ref{FHTrend}.

\subsection{Spectral slope}
\label{slope}
Figure~\ref{optspec} shows that in the optical, the spectral slopes of our two targets are quite similar to each other and resemble the normal early-T dwarf slope. We used the least-squares method to compare the two spectra with the entire T dwarf standard grid provided by \citep{burgasser2003dT_optical}.  
We used the wavelength range from $8000\AA$ to $9250\AA$.  
We found that WISE0004 matches T2 best, and WISE0301 
matches T1. This is in agreement with the spectral types 
estimated from the NIR spectra for WISE0004 (T2) and 
WISE0301 (T1) by \citet{greco2019neowise}. Although 
\citet{williambest2020_L0-8UKIRTparallax}  estimate a type 
T4 for WISE0004.

\subsection{Hydrides}
WISE0301 shows a CrH bandhead starting at $8611$ \AA\ that is just 
like the T1 dwarf standard, but WISE0004 does not show it. 
This absorption feature disappears from spectral classes T1 to T5. 
These effects are illustrated in the lower left 
panel of Figure~\ref{component}.  

Both targets have deeper FeH  absorption at $9896$ \AA\ than 
the T1 standard and are similar to the T5 standard. This 
can be seen in the lower right panel of Figure~\ref{component}. 
Neither object shows strong CrH at $9969$ \AA\ or FeH at 
$8692$ \AA\, just like their solar-metallicity counterparts.

\subsection{Alkali atomic lines}

The Na I and K I resonance doublets are known to be 
extremely broad in ultracool dwarfs \citep{martin1999Lclassification}. 
The red wing of the Na I feature in our two targets 
shapes the optical spectra from $5900$ \AA\ to $7500$ \AA\, 
as can be seen in the upper logarithmic-scale plot 
in Figure~\ref{optspec} and the upper right linear-scale plot
 in Figure~\ref{component}.  In the same figures, 
the strong K I resonance doublet at $7665$ \AA\ and
 $7699$ \AA\ forms V-shaped notches in both targets. 
These are as deep and wide as that of the T5 standard.

The extended absorption due to the Na I and K I resonance 
features render the signal-to-noise ratio (S/N) in the 
continuum  very low in our spectra shortwards $8000 
\AA$ and prevent us from detecting or setting any significant 
limits to the presence of the Li I resonance doublet at 
$6708$ \AA\ (a trademark of substellar status) or the Rb I 
resonance doublet at $7800$ \AA\ and $7948$ \AA. 


The Cs I resonance doublet at $8521.1$ \AA\ and 
$8943.5$ \AA\ is prominent. We used the IRAF programmes 
\textit{rspec}  and  \textit{splot} to convert the 
\textit{Pypeit} fits table to fits image and then to 
measure the equivalent width (EW) of these two lines. 
The measurement of the second line is tricky because 
it is located within the CH$_4$ absorption band. The 
first line of WISE0004 has an EW of $8.9$ \AA\ and 
of WISE0301 has an EW of $8.4$ \AA\; the second 
line of WISE0004 has an EW of $8.4$ \AA\ and of WISE0301
 has an EW of $7.4$ \AA.  \citet{burgasser2003dT_optical}
reported that for early and mid-T dwarfs both of the
caesium lines have an EW of about 7 to 9 $\AA$, and  
\citet{lodieu2015luhman16xshooter} reported that 
the EW of the caesium $8521.1$ \AA\ line of normal T dwarfs
 peaks in early-T dwarfs is about $8$ \AA.  The 
two T subdwarfs have EWs comparable to the EWs of 
normal T dwarfs. 

\subsection{Water bands}
\label{waterbands}
\begin{figure}[H]
    \centering
    \includegraphics[width=0.5\textwidth]{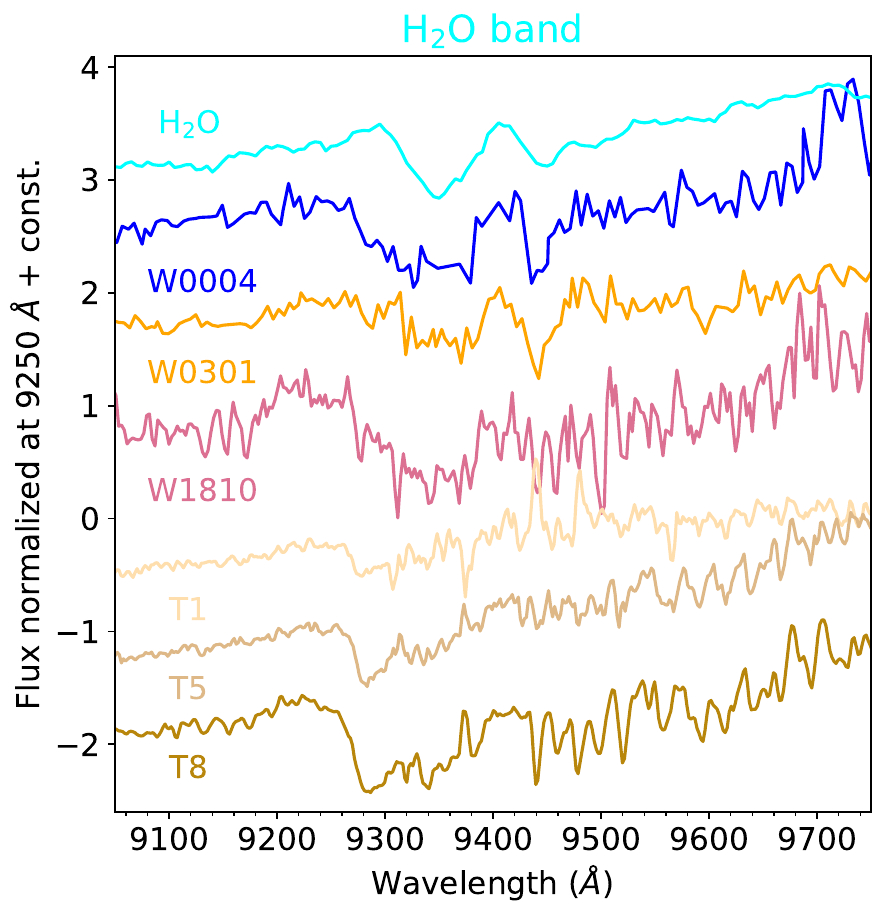}
    \caption{The GTC OSIRIS spectra of WISE0004, 
WISE0301 and WISE1810 together with three T dwarf 
templates (Keck LRIS spectra) in the water band spectral 
region and a laboratory spectrum of pure H$_2$O gas. }
    \label{lab}
\end{figure}

We compared our GTC OSIRIS T subdwarf spectra and WISE1810 spectra with normal T dwarf standards and with a laboratory spectrum of pure H$_2$O gas \citep{martin2021ch4nh3}. 
The normal T dwarfs show a H$_2$O bandhead that reaches the deepest level at the beginning of the band at $9280$ \AA\ and  attenuates for longer wavelengths. The T subdwarfs have the deepest absorption feature at $9350$ \AA\, like the laboratory water spectrum. The two T subdwarfs also have a second absorption feature centred at $9450$ \AA\ that is also seen in the laboratory spectrum but not in the normal T dwarfs. See the comparison in Figure~\ref{lab}. 

To explain the similarity of the optical water bands in the T subdwarfs and the laboratory spectrum, and the difference from the normal T dwarfs, we speculate that the physical conditions in metal-poor atmospheres may favour the condensation of pure water molecules that interact more among themselves and less with other molecules and with atoms than in the case of metal-rich atmospheres.  

\section{Metallicity Gradient}
\label{FHTrend}

\begin{figure*}
\centering   \includegraphics[width=\textwidth]{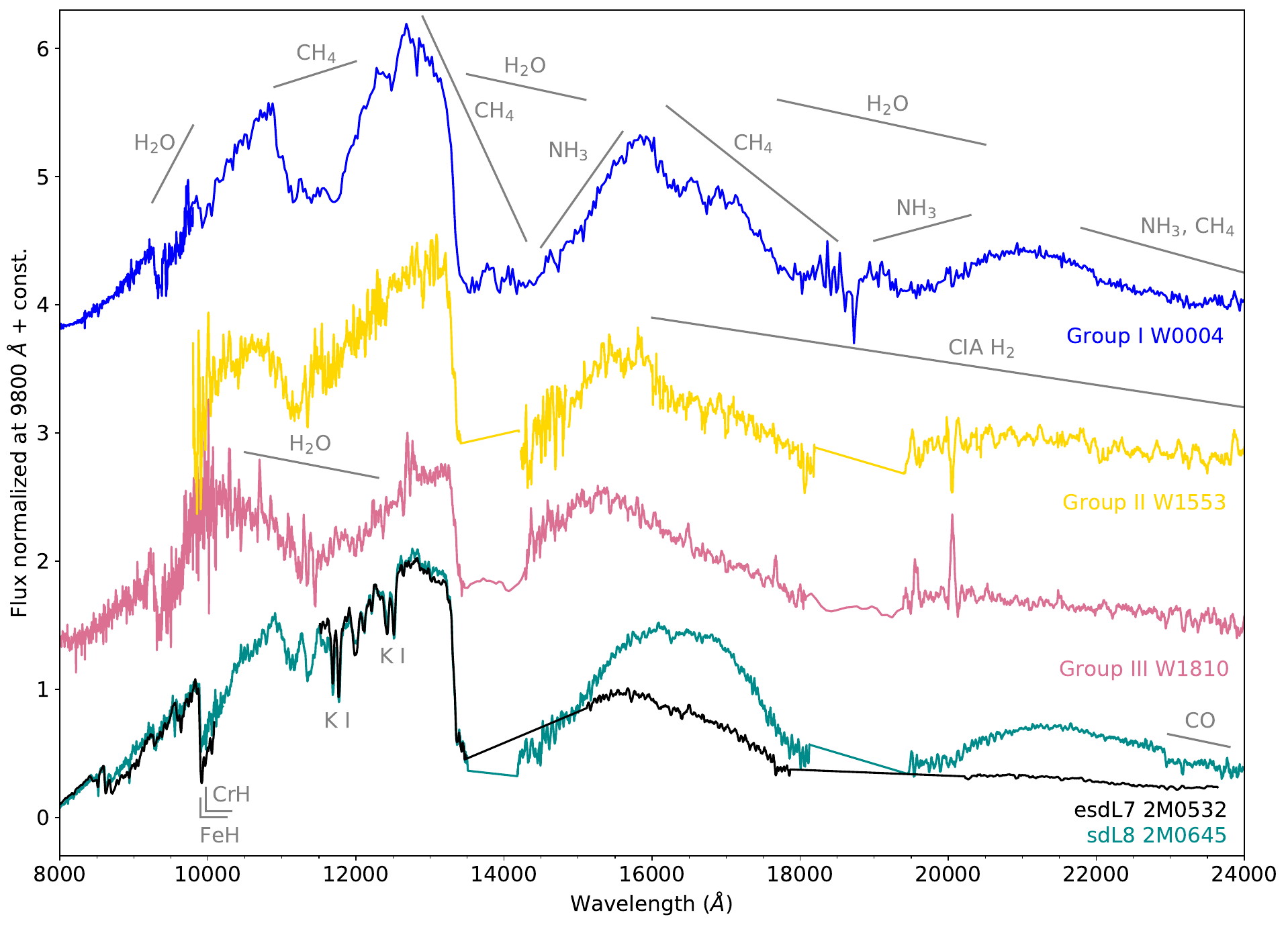}
    \caption{Normalized optical and NIR spectra (from the
$z$- to  the $K$-band) of five representative metal-poor 
ultracool dwarfs: the sdT2 WISE0004 from Group I in the 
colour--colour diagram shown in Figure~\ref{colourcolour} 
(the optical part from this research and the NIR part 
from \citet{greco2019neowise}); the Z class prototype 
WISE1810 from Group III (the optical part from \citet{lodieu2022W1810} 
and the NIR part from \citet{Schneider2020W0414_W1810}); 
the sdT4 WISE1553 from Group II in between Group I and 
Group III (only NIR spectrum from \citet{meisner2021esdT}); 
the sdL8 2MASS0645 \citep{zhangzenghua2018subL}, and the 
esdL7 2MASS0532 \citep{burgasser2003esdL7_2M0532}. The 
major absorption bands and atomic lines are marked.}
    \label{esdL}
\end{figure*}


Figure~\ref{esdL} illustrates how the optical and NIR 
spectra change in Group I, II, III. For the Group I object WISE0004, the optical and the NIR part are from this study and NASA IRTF \citep{greco2019neowise}, respectively. For the Group II object WISE1553 there is only the NIR spectrum from KECK NIRES 
\citep{meisner2021esdT}. We smoothed its spectrum with
 a window size of 21 to match the spectral resolution 
of other spectra. The optical spectrum of Group III object WISE1810 is from GTC OSIRIS \citep{lodieu2022W1810}, and the 
NIR part is from Palomar/TripleSpec \citep{Schneider2020W0414_W1810}
.

As can be seen in Figure~\ref{esdL}, Group I keeps the three NIR flux peaks 
of normal T dwarfs at 1.25, 1.60 and 2.10 $\mu$m 
\citep{burgasser2002dT_spec_classification} at the same 
positions, as well as the peak at 1.08 $\mu$m in the 
optical far red.  Group III,  has the optical far red peak at 1.00 $\mu$m because of very broad triangular shaped  H$_2$O absorption that dominates in the $J$-band window.  The NIR flux peaks are located at 1.3 $\mu$m, owing to the lack of CH$_4$ absorption in the $J$-band, and 1.52 $\mu$m, possibly because of the lack of NH$_3$. The flux peak at 2.0 $\mu$m is flattened by the CIA of H$_2$. 

The CH$_4$ absorption in the NIR is the trademark of 
T dwarfs and is shown by both T subdwarfs in Group I and II. Methane is what distinguishes T dwarfs from L dwarfs; in fact T dwarfs are also known as methane dwarfs. However, the CH$_4$ features are very weak or completely missing in the Group III objects. The disappearance of methane is a clear sign of very low metallicity.

\subsection{The $z-W1$ colour as new metallicity indicator}
\label{z-w1}

There is a gradient of the spectral morphology between the groups: From Group I to Group III, the $J$-band H$_2$O absorption is getting stronger; the peak around 1.3 $\mu$m moves to the red; the $H$-band CH$_4$ absorption is getting weaker; the weaker NH$_3$ absorption makes the third peak and the fourth peak moves blueward; and the $H$- and $K$-bands get more suppressed by the CIA of H$_2$.

The metallicity estimation of T subdwarfs relies strongly on atmospheric modelling and it is very hard to consider developing at the present time a quantitative relationship between the physical observables and the metallicity values with such a small sample. The spectral morphology, however, can show the metallicity trend. Constructing a qualitative metallicity indicator based on this trend is feasible.

Group III objects have a estimated/approximate metallicity between $-$1.5 and $-$1.0 dex \citep{Schneider2020W0414_W1810,lodieu2022W1810}. \citet{meisner2021esdT} reported that WISE1553 has a metallicity [Fe/H] $\approx-0.5$ according to the NIR spectroscopy and the PHOENIX atmospheric models \citep{hauschildt1999expand_atm_numerical,allard2013atm_model} extending to low metallicity \citep{gerasimov2020pheonix_metalpoor}, but [Fe/H] $\lesssim-1.5$ according to the broad-band photometry and the LOWZ model of the author. Group I was classified as a T subdwarf but the exact metallicity is unknown. According to the most similarity between the normal T dwarf spectra and Group I spectra, we imply that they have metallicity close to the normal T dwarfs, i.e., $-0.5<$ [Fe/H] $<0$.

Overall, we observe a general decreasing trend of metallicity from 0 to $-$1.5 dex from Group I to Group III that coincide with the increase of the $z-W1$ colour from 5 mag to 7 mag. Therefore we argue that the robustness of the $z-W1$ colour might be a qualitative metallicity indicator, for ultracool dwarfs with $W1-W2$ colour redder than 0.8 mag, which have temperatures lower than that of normal L dwarfs. 

\citet{cushing2005IR_MLT} showed there is a strong methane absorption in the MIR from 3.0 $\mu$m to 3.8 $\mu$m (fully covered by the $W1$-band) when the T dwarf sequence starts. The reddening of the $z-W1$ colour from Group I to Group III can be attributed to the attenuation of the major methane absorption in the $W1$-band
We also expect that there should be a limit for this indicator when the metallicity becomes too low and the $W1$ band is no longer affected by methane. Further MIR observations are needed to support this hypothesis and determine the lowest metallicity that this indicator can still be valid.

\subsection{Extreme T subdwarf?}
Two objects in Group III, WISE1810 and WISE0414 were tentatively designated as esdT0.0$\pm$1.0 \citep{Schneider2020W0414_W1810}. However, the lack of methane absorption features at the NIR wavelengths raises the question on the spectral classification of these objects which might require a revision once JWST MIR become available.

\citet{Schneider2020W0414_W1810} also noticed that the 
CIA of hydrogen shapes the spectra of WISE1810 and WISE0414 in a way that resembles the extreme L subdwarfs in $H$ and $K$ bands, indicating a similar metallicity to the extreme L subdwarf. Indeed, \citet{lodieu2022W1810} assigned a metallicity [Fe/H] $\approx-1.5$ to WISE1810, which falls in the metallicity range of the extreme L subdwarfs, [Fe/H] from $-1.0$ down to $-1.7$ \citep{zhang2017six_sdL_classification}. So are Group III objects in fact late-L subdwarfs? To answer 
this, we show another comparison in the lower part of 
Figure~\ref{esdL}. The sdL8 
2MASS\,J06453153$-$6646120 represents an L subdwarf 
(both the optical and the NIR part were taken by 
\citet{zhangzenghua2018subL} using X-shooter on the VLT, 
Chile), together with the esdL7 2MASS\,J05325346+8246465 
from \citet{burgasser2003esdL7_2M0532} using LRIS and 
NIRSPEC on the Keck I and II. 

First of all, Group III objects have a redder $W1-W2$ colour, 
not mixing with any L dwarfs or L subdwarfs in the 
$W1-W2$ vs.\ $z-W1$ colour--colour diagram. Second, The 
hydride bands, especially FeH, become stronger when 
metallicity decreases \citep{burgasser2007sdLoptical,zhangzenghua2018subL}, 
as seen from the comparison between the sdL8 and esdL7 
in Figure~\ref{esdL}. However, Group III objects do not show 
any signs of hydride in their optical or NIR spectra. 
Thirdly, the L subdwarfs and extreme L subdwarfs show 
very little H$_2$O absorption, but Group III objects have very 
strong H$_2$O bands in both  the $z$ and the $J$-bands. Group III objects also do not show the strong potassium K I lines in the 
$J$-band, which appear strong in both late-L 
subdwarf and extreme subdwarf spectra. 


\section{Conclusions}
\label{conclusion}
The photometric and spectroscopic characterization of metal-poor BDs widens the parameter space over which substellar mass objects are investigated. We used GTC OSIRIS to obtain $z$-band photometry for ten T subdwarf candidates, and optical spectroscopy for two of them, namely WISE0004 and WISE0301.  We downloaded $z$-band photometry of WISE0414 from the DES database.

We noticed that our targets segregated into three subgroups 
in the $W1-W2$ vs.\ $z-W1$ colour--colour diagram. Group I 
mixed with the T dwarf sequence from the literature; Group 
III is composed of only three objects (WISE1810, WISE0414 and 
WISE2217) and has extraordinarily red $z-W1$ colours. 
Group II is located between the other two groups. 

We examined the optical spectra of two early-T subdwarfs in 
Group I (WISE0004 and WISE0301). Their optical spectra appear 
similar to those of early-T dwarfs that are dominated 
by alkali metals and hydrides. We inferred that these two 
early-T subdwarfs do not have very low metallicity. 
Nevertheless, they have a deeper water absorption feature in 
the optical. This water absorption feature in T subdwarf 
spectra is more similar to the water transmission absorption 
spectra obtained in the laboratory compared with normal T 
dwarf spectra. 

Based on the peculiar position of these T subdwarf candidates in the colour--colour diagram, we compared the spectra of the early-T subdwarf WISE0004 in Group I, mid-T subdwarf WISE1553 in Group II, WISE1810 in Group III, the late-L subdwarf 2MASS0645 and the late-L extreme subdwarf 2MASS0532. We conclude that the $z-W1$ colour is qualitatively a good metallicity indicator for objects with temperatures comparable to that of T dwarfs.

The spectroscopy of the late-type objects in Group II, WISE0422, 
is needed in the future to test the robustness of the $z-W1$ colour as a metallicity indicator for cooler objects; as well as the spectroscopy of the third object in Group III, WISE2217, to confirm its metal-poor nature, although it will be extremely challenging ($J$\,=\,20.66 mag).

The spectral classification of Group III objects might require a revision because of the weak or absent methane absorption at NIR wavelengths, a key molecular feature to classify T-type dwarfs.  We anticipate that new JWST MIR observations will bring new insight on the spectral classification of metal-poor BDs. An option could be using new letters to classify spectra that are different to those previously classified with the letters L, T and Y. We note that the letters H and Z are still available.

More objects of these three groups, and those filling the gaps between them need to be discovered to improve the statistical significance of the connection between the $z-W1$ colour and metallicity.  We also need more accurate metallicity measurements of these objects to be able to establish this relationship. To achieve this is promising in the near future. A quantum leap is expected in the numbers of substellar objects that can be identified with the advent of deep large area surveys such as the Euclid space mission and Legacy Survey of Space and Time (LSST) at the Vera Rubin Observatory. Thousands of T dwarf slitless low-resolution NIR spectra are expected from the Euclid surveys \citep{martin2021ch4nh3}, and LSST will become a powerful search engine for Group III objects and those even more metal-poor, expanding more than 1000 times of the space volume in the solar vicinity that we have explored so far.\footnote{Here we assumed LSST coadded 5-$\sigma$ $z$-band limit to be 25.6 mag, and considering the fact that WISE1810 has 20.7 mag in $z_{AB}$. LSST $z$-band filter is similar to Pan-STARRS's. About the LSST depth calculation see \url{https://smtn-002.lsst.io}.}

\begin{acknowledgements}
Funding for this research was provided by the European 
Union (ERC, SUBSTELLAR, project number 101054354) and the 
Agencia Estatal de Investigación del Ministerio de Ciencia 
e Innovación (AEI-MCINN) under grant PID2019-109522GB-C53\@.   
Based on observations made with the Gran Telescopio Canarias 
(GTC), installed at the Spanish Observatorio del Roque de 
los Muchachos of the Instituto de Astrofísica de Canarias, 
on the island of La Palma. This work is (partly) based on 
data obtained with the instrument OSIRIS, built by a 
Consortium led by the Instituto de Astrofísica de Canarias 
in collaboration with the Instituto de Astronomía of the 
Universidad Autónoma de México. OSIRIS was funded by GRANTECAN 
and the National Plan of Astronomy and Astrophysics of the 
Spanish Government. This research has made use of data provided by 
Astrometry.net. This work has used the Pan-STARRS1 Surveys 
(PS1) and the PS1 public science archive that have been made 
possible through contributions by the Institute for Astronomy, 
the University of Hawaii, the Pan-STARRS Project Office, the 
Max-Planck Society and its participating institutes, the Max 
Planck Institute for Astronomy, Heidelberg and the Max Planck 
Institute for Extraterrestrial Physics, Garching, The Johns 
Hopkins University, Durham University, the University of Edinburgh, 
the Queen's University Belfast, the Harvard-Smithsonian Center 
for Astrophysics, the Las Cumbres Observatory Global Telescope 
Network Incorporated, the National Central University of 
Taiwan, the Space Telescope Science Institute, the National 
Aeronautics and Space Administration under Grant No. NNX08AR22G 
issued through the Planetary Science Division of the NASA Science
 Mission Directorate, the National Science Foundation Grant No. 
AST-1238877, the University of Maryland, Eotvos Lorand University 
(ELTE), the Los Alamos National Laboratory, and the Gordon and 
Betty Moore Foundation. This publication makes use of data products from the Wide-field Infrared Survey Explorer, which is a joint project of the University of California, Los Angeles, and the Jet Propulsion Laboratory/California Institute of Technology, funded by the National Aeronautics and Space Administration. This project used public archival data from the Dark Energy Survey (DES). Funding for the DES Projects has been provided by the U.S. Department of Energy, the U.S. National Science Foundation, the Ministry of Science and Education of Spain, the Science and Technology Facilities Council of the United Kingdom, the Higher Education Funding Council for England, the National Center for Supercomputing Applications at the University of Illinois at Urbana–Champaign, the Kavli Institute of Cosmological Physics at the University of Chicago, the Center for Cosmology and Astro-Particle Physics at the Ohio State University, the Mitchell Institute for Fundamental Physics and Astronomy at Texas A\&M University, Financiadora de Estudos e Projetos, Fundação Carlos Chagas Filho de Amparo à Pesquisa do Estado do Rio de Janeiro, Conselho Nacional de Desenvolvimento Científico e Tecnológico and the Ministério da Ciência, Tecnologia e Inovação, the Deutsche Forschungsgemeinschaft and the Collaborating Institutions in the Dark Energy Survey.
The Collaborating Institutions are Argonne National Laboratory, the University of California at Santa Cruz, the University of Cambridge, Centro de Investigaciones Enérgeticas, Medioambientales y Tecnológicas–Madrid, the University of Chicago, University College London, the DES-Brazil Consortium, the University of Edinburgh, the Eidgenössische Technische Hochschule (ETH) Zürich, Fermi National Accelerator Laboratory, the University of Illinois at Urbana-Champaign, the Institut de Ciències de l’Espai (IEEC/CSIC), the Institut de Física d’Altes Energies, Lawrence Berkeley National Laboratory, the Ludwig-Maximilians Universität München and the associated Excellence Cluster Universe, the University of Michigan, the National Optical Astronomy Observatory, the University of Nottingham, The Ohio State University, the OzDES Membership Consortium, the University of Pennsylvania, the University of Portsmouth, SLAC National Accelerator Laboratory, Stanford University, the University of Sussex, and Texas A\&M University.
Based in part on observations at Cerro Tololo Inter-American Observatory, National Optical Astronomy Observatory, which is operated by the Association of Universities for Research in Astronomy (AURA) under a cooperative agreement with the National Science Foundation. 
This work made use of Astropy:\footnote{http://www.astropy.org} a community-developed core Python package and an ecosystem of tools and resources for astronomy \citep{astropy2013, astropy2018, astropy2022}. We appreciate the referee report for providing useful and insightful comments.
\end{acknowledgements}

%
%

%

\bibliographystyle{aa} 
\bibliography{bibliography.bib} 

\begin{thebibliography}{73}
\expandafter\ifx\csname natexlab\endcsname\relax\def\natexlab#1{#1}\fi

\bibitem[{{Abbott} {et~al.}(2018){Abbott}, {Abdalla}, {Allam}, {Amara},
  {Annis}, {Asorey}, {Avila}, {Ballester}, {Banerji}, {Barkhouse}, {Baruah},
  {Baumer}, {Bechtol}, {Becker}, {Benoit-L{\'e}vy}, {Bernstein}, {Bertin},
  {Blazek}, {Bocquet}, {Brooks}, {Brout}, {Buckley-Geer}, {Burke}, {Busti},
  {Campisano}, {Cardiel-Sas}, {Carnero Rosell}, {Carrasco Kind}, {Carretero},
  {Castander}, {Cawthon}, {Chang}, {Chen}, {Conselice}, {Costa}, {Crocce},
  {Cunha}, {D'Andrea}, {da Costa}, {Das}, {Daues}, {Davis}, {Davis}, {De
  Vicente}, {DePoy}, {DeRose}, {Desai}, {Diehl}, {Dietrich}, {Dodelson},
  {Doel}, {Drlica-Wagner}, {Eifler}, {Elliott}, {Evrard}, {Farahi}, {Fausti
  Neto}, {Fernandez}, {Finley}, {Flaugher}, {Foley}, {Fosalba}, {Friedel},
  {Frieman}, {Garc{\'\i}a-Bellido}, {Gaztanaga}, {Gerdes}, {Giannantonio},
  {Gill}, {Glazebrook}, {Goldstein}, {Gower}, {Gruen}, {Gruendl}, {Gschwend},
  {Gupta}, {Gutierrez}, {Hamilton}, {Hartley}, {Hinton}, {Hislop}, {Hollowood},
  {Honscheid}, {Hoyle}, {Huterer}, {Jain}, {James}, {Jeltema}, {Johnson},
  {Johnson}, {Kacprzak}, {Kent}, {Khullar}, {Klein}, {Kovacs}, {Koziol},
  {Krause}, {Kremin}, {Kron}, {Kuehn}, {Kuhlmann}, {Kuropatkin}, {Lahav},
  {Lasker}, {Li}, {Li}, {Liddle}, {Lima}, {Lin}, {L{\'o}pez-Reyes}, {MacCrann},
  {Maia}, {Maloney}, {Manera}, {March}, {Marriner}, {Marshall}, {Martini},
  {McClintock}, {McKay}, {McMahon}, {Melchior}, {Menanteau}, {Miller},
  {Miquel}, {Mohr}, {Morganson}, {Mould}, {Neilsen}, {Nichol}, {Nogueira},
  {Nord}, {Nugent}, {Nunes}, {Ogando}, {Old}, {Pace}, {Palmese},
  {Paz-Chinch{\'o}n}, {Peiris}, {Percival}, {Petravick}, {Plazas}, {Poh},
  {Pond}, {Porredon}, {Pujol}, {Refregier}, {Reil}, {Ricker}, {Rollins},
  {Romer}, {Roodman}, {Rooney}, {Ross}, {Rykoff}, {Sako}, {Sanchez}, {Sanchez},
  {Santiago}, {Saro}, {Scarpine}, {Scolnic}, {Serrano}, {Sevilla-Noarbe},
  {Sheldon}, {Shipp}, {Silveira}, {Smith}, {Smith}, {Smith}, {Soares-Santos},
  {Sobreira}, {Song}, {Stebbins}, {Suchyta}, {Sullivan}, {Swanson}, {Tarle},
  {Thaler}, {Thomas}, {Thomas}, {Troxel}, {Tucker}, {Vikram}, {Vivas},
  {Walker}, {Wechsler}, {Weller}, {Wester}, {Wolf}, {Wu}, {Yanny}, {Zenteno},
  {Zhang}, {Zuntz}, {DES Collaboration}, {Juneau}, {Fitzpatrick}, {Nikutta},
  {Nidever}, {Olsen}, {Scott}, \& {NOAO Data Lab}}]{abbott2019DESdr1}
{Abbott}, T.~M.~C., {Abdalla}, F.~B., {Allam}, S., {et~al.} 2018, \apjs, 239,
  18

\bibitem[{{Allard} {et~al.}(2013){Allard}, {Homeier}, {Freytag},
  {Schaffenberger}, \& {Rajpurohit}}]{allard2013atm_model}
{Allard}, F., {Homeier}, D., {Freytag}, B., {Schaffenberger}, W., \&
  {Rajpurohit}, A.~S. 2013, Memorie della Societa Astronomica Italiana
  Supplementi, 24, 128

\bibitem[{{Astropy Collaboration} {et~al.}(2022){Astropy Collaboration},
  {Price-Whelan}, {Lim}, {Earl}, {Starkman}, {Bradley}, {Shupe}, {Patil},
  {Corrales}, {Brasseur}, {N{\"o}the}, {Donath}, {Tollerud}, {Morris},
  {Ginsburg}, {Vaher}, {Weaver}, {Tocknell}, {Jamieson}, {van Kerkwijk},
  {Robitaille}, {Merry}, {Bachetti}, {G{\"u}nther}, {Aldcroft},
  {Alvarado-Montes}, {Archibald}, {B{\'o}di}, {Bapat}, {Barentsen},
  {Baz{\'a}n}, {Biswas}, {Boquien}, {Burke}, {Cara}, {Cara}, {Conroy},
  {Conseil}, {Craig}, {Cross}, {Cruz}, {D'Eugenio}, {Dencheva}, {Devillepoix},
  {Dietrich}, {Eigenbrot}, {Erben}, {Ferreira}, {Foreman-Mackey}, {Fox},
  {Freij}, {Garg}, {Geda}, {Glattly}, {Gondhalekar}, {Gordon}, {Grant},
  {Greenfield}, {Groener}, {Guest}, {Gurovich}, {Handberg}, {Hart},
  {Hatfield-Dodds}, {Homeier}, {Hosseinzadeh}, {Jenness}, {Jones}, {Joseph},
  {Kalmbach}, {Karamehmetoglu}, {Ka{\l}uszy{\'n}ski}, {Kelley}, {Kern},
  {Kerzendorf}, {Koch}, {Kulumani}, {Lee}, {Ly}, {Ma}, {MacBride}, {Maljaars},
  {Muna}, {Murphy}, {Norman}, {O'Steen}, {Oman}, {Pacifici}, {Pascual},
  {Pascual-Granado}, {Patil}, {Perren}, {Pickering}, {Rastogi}, {Roulston},
  {Ryan}, {Rykoff}, {Sabater}, {Sakurikar}, {Salgado}, {Sanghi}, {Saunders},
  {Savchenko}, {Schwardt}, {Seifert-Eckert}, {Shih}, {Jain}, {Shukla}, {Sick},
  {Simpson}, {Singanamalla}, {Singer}, {Singhal}, {Sinha}, {Sip{\H{o}}cz},
  {Spitler}, {Stansby}, {Streicher}, {{\v{S}}umak}, {Swinbank}, {Taranu},
  {Tewary}, {Tremblay}, {de Val-Borro}, {Van Kooten}, {Vasovi{\'c}}, {Verma},
  {de Miranda Cardoso}, {Williams}, {Wilson}, {Winkel}, {Wood-Vasey}, {Xue},
  {Yoachim}, {Zhang}, {Zonca}, \& {Astropy Project Contributors}}]{astropy2022}
{Astropy Collaboration}, {Price-Whelan}, A.~M., {Lim}, P.~L., {et~al.} 2022,
  \apj, 935, 167

\bibitem[{{Astropy Collaboration} {et~al.}(2018){Astropy Collaboration},
  {Price-Whelan}, {Sip{\H{o}}cz}, {G{\"u}nther}, {Lim}, {Crawford}, {Conseil},
  {Shupe}, {Craig}, {Dencheva}, {Ginsburg}, {VanderPlas}, {Bradley},
  {P{\'e}rez-Su{\'a}rez}, {de Val-Borro}, {Aldcroft}, {Cruz}, {Robitaille},
  {Tollerud}, {Ardelean}, {Babej}, {Bach}, {Bachetti}, {Bakanov}, {Bamford},
  {Barentsen}, {Barmby}, {Baumbach}, {Berry}, {Biscani}, {Boquien}, {Bostroem},
  {Bouma}, {Brammer}, {Bray}, {Breytenbach}, {Buddelmeijer}, {Burke},
  {Calderone}, {Cano Rodr{\'\i}guez}, {Cara}, {Cardoso}, {Cheedella}, {Copin},
  {Corrales}, {Crichton}, {D'Avella}, {Deil}, {Depagne}, {Dietrich}, {Donath},
  {Droettboom}, {Earl}, {Erben}, {Fabbro}, {Ferreira}, {Finethy}, {Fox},
  {Garrison}, {Gibbons}, {Goldstein}, {Gommers}, {Greco}, {Greenfield},
  {Groener}, {Grollier}, {Hagen}, {Hirst}, {Homeier}, {Horton}, {Hosseinzadeh},
  {Hu}, {Hunkeler}, {Ivezi{\'c}}, {Jain}, {Jenness}, {Kanarek}, {Kendrew},
  {Kern}, {Kerzendorf}, {Khvalko}, {King}, {Kirkby}, {Kulkarni}, {Kumar},
  {Lee}, {Lenz}, {Littlefair}, {Ma}, {Macleod}, {Mastropietro}, {McCully},
  {Montagnac}, {Morris}, {Mueller}, {Mumford}, {Muna}, {Murphy}, {Nelson},
  {Nguyen}, {Ninan}, {N{\"o}the}, {Ogaz}, {Oh}, {Parejko}, {Parley}, {Pascual},
  {Patil}, {Patil}, {Plunkett}, {Prochaska}, {Rastogi}, {Reddy Janga},
  {Sabater}, {Sakurikar}, {Seifert}, {Sherbert}, {Sherwood-Taylor}, {Shih},
  {Sick}, {Silbiger}, {Singanamalla}, {Singer}, {Sladen}, {Sooley},
  {Sornarajah}, {Streicher}, {Teuben}, {Thomas}, {Tremblay}, {Turner},
  {Terr{\'o}n}, {van Kerkwijk}, {de la Vega}, {Watkins}, {Weaver}, {Whitmore},
  {Woillez}, {Zabalza}, \& {Astropy Contributors}}]{astropy2018}
{Astropy Collaboration}, {Price-Whelan}, A.~M., {Sip{\H{o}}cz}, B.~M., {et~al.}
  2018, \aj, 156, 123

\bibitem[{{Astropy Collaboration} {et~al.}(2013){Astropy Collaboration},
  {Robitaille}, {Tollerud}, {Greenfield}, {Droettboom}, {Bray}, {Aldcroft},
  {Davis}, {Ginsburg}, {Price-Whelan}, {Kerzendorf}, {Conley}, {Crighton},
  {Barbary}, {Muna}, {Ferguson}, {Grollier}, {Parikh}, {Nair}, {Unther},
  {Deil}, {Woillez}, {Conseil}, {Kramer}, {Turner}, {Singer}, {Fox}, {Weaver},
  {Zabalza}, {Edwards}, {Azalee Bostroem}, {Burke}, {Casey}, {Crawford},
  {Dencheva}, {Ely}, {Jenness}, {Labrie}, {Lim}, {Pierfederici}, {Pontzen},
  {Ptak}, {Refsdal}, {Servillat}, \& {Streicher}}]{astropy2013}
{Astropy Collaboration}, {Robitaille}, T.~P., {Tollerud}, E.~J., {et~al.} 2013,
  \aap, 558, A33

\bibitem[{Baraffe {et~al.}(1995)Baraffe, Chabrier, Allard, \&
  Hauschildt}]{baraffe1995evolutionBD}
Baraffe, I., Chabrier, G., Allard, F., \& Hauschildt, P. 1995, The
  Astrophysical Journal, 446, L35

\bibitem[{Basri {et~al.}(1996)Basri, Marcy, \& Graham}]{basri1996lithium}
Basri, G., Marcy, G.~W., \& Graham, J.~R. 1996, The Astrophysical Journal, 458,
  600

\bibitem[{{Best} {et~al.}(2020){Best}, {Liu}, {Magnier}, \&
  {Dupuy}}]{williambest2020_L0-8UKIRTparallax}
{Best}, W. M.~J., {Liu}, M.~C., {Magnier}, E.~A., \& {Dupuy}, T.~J. 2020, \aj,
  159, 257

\bibitem[{{Best} {et~al.}(2018){Best}, {Magnier}, {Liu}, {Aller}, {Zhang},
  {Burgett}, {Chambers}, {Draper}, {Flewelling}, {Kaiser}, {Kudritzki},
  {Metcalfe}, {Tonry}, {Wainscoat}, \& {Waters}}]{best2018PS1_3pi}
{Best}, W. M.~J., {Magnier}, E.~A., {Liu}, M.~C., {et~al.} 2018, \apjs, 234, 1

\bibitem[{{Burgasser} {et~al.}(2007){Burgasser}, {Cruz}, \&
  {Kirkpatrick}}]{burgasser2007sdLoptical}
{Burgasser}, A.~J., {Cruz}, K.~L., \& {Kirkpatrick}, J.~D. 2007, \apj, 657, 494

\bibitem[{{Burgasser} {et~al.}(2002){Burgasser}, {Kirkpatrick}, {Brown},
  {Reid}, {Burrows}, {Liebert}, {Matthews}, {Gizis}, {Dahn}, {Monet}, {Cutri},
  \& {Skrutskie}}]{burgasser2002dT_spec_classification}
{Burgasser}, A.~J., {Kirkpatrick}, J.~D., {Brown}, M.~E., {et~al.} 2002, \apj,
  564, 421

\bibitem[{{Burgasser} {et~al.}(2003{\natexlab{a}}){Burgasser}, {Kirkpatrick},
  {Burrows}, {Liebert}, {Reid}, {Gizis}, {McGovern}, {Prato}, \&
  {McLean}}]{burgasser2003esdL7_2M0532}
{Burgasser}, A.~J., {Kirkpatrick}, J.~D., {Burrows}, A., {et~al.}
  2003{\natexlab{a}}, \apj, 592, 1186

\bibitem[{{Burgasser} {et~al.}(2003{\natexlab{b}}){Burgasser}, {Kirkpatrick},
  {Liebert}, \& {Burrows}}]{burgasser2003dT_optical}
{Burgasser}, A.~J., {Kirkpatrick}, J.~D., {Liebert}, J., \& {Burrows}, A.
  2003{\natexlab{b}}, \apj, 594, 510

\bibitem[{{Burningham} {et~al.}(2010){Burningham}, {Pinfield}, {Lucas},
  {Leggett}, {Deacon}, {Tamura}, {Tinney}, {Lodieu}, {Zhang}, {Huelamo},
  {Jones}, {Murray}, {Mortlock}, {Patel}, {Barrado Y Navascu{\'e}s}, {Zapatero
  Osorio}, {Ishii}, {Kuzuhara}, \& {Smart}}]{Burningham2010_47dT}
{Burningham}, B., {Pinfield}, D.~J., {Lucas}, P.~W., {et~al.} 2010, \mnras,
  406, 1885

\bibitem[{{Burningham} {et~al.}(2014){Burningham}, {Smith}, {Cardoso}, {Lucas},
  {Burgasser}, {Jones}, \& {Smart}}]{burningham2014T6.5}
{Burningham}, B., {Smith}, L., {Cardoso}, C.~V., {et~al.} 2014, \mnras, 440,
  359

\bibitem[{Burrows \& Liebert(1993)}]{burrows1993scienceBD}
Burrows, A. \& Liebert, J. 1993, Reviews of Modern Physics, 65, 301

\bibitem[{Cepa {et~al.}(2000)Cepa, Aguiar-Gonzalez, Gonzalez-Escalera,
  Gonzalez-Serrano, Joven-Alvarez, Cano, Rasilla, Rodriguez-Ramos,
  Gonz{\'a}lez, Duenas, {et~al.}}]{cepa2000osiris}
Cepa, J., Aguiar-Gonzalez, M., Gonzalez-Escalera, V., {et~al.} 2000, in Optical
  and IR Telescope Instrumentation and Detectors, Vol. 4008, SPIE, 623--631

\bibitem[{Chabrier {et~al.}(2000)Chabrier, Baraffe, Allard, \&
  Hauschildt}]{chabrier2000deuterium}
Chabrier, G., Baraffe, I., Allard, F., \& Hauschildt, P. 2000, The
  Astrophysical Journal, 542, L119

\bibitem[{{Chambers} {et~al.}(2016){Chambers}, {Magnier}, {Metcalfe},
  {Flewelling}, {Huber}, {Waters}, {Denneau}, {Draper}, {Farrow}, {Finkbeiner},
  {Holmberg}, {Koppenhoefer}, {Price}, {Rest}, {Saglia}, {Schlafly}, {Smartt},
  {Sweeney}, {Wainscoat}, {Burgett}, {Chastel}, {Grav}, {Heasley}, {Hodapp},
  {Jedicke}, {Kaiser}, {Kudritzki}, {Luppino}, {Lupton}, {Monet}, {Morgan},
  {Onaka}, {Shiao}, {Stubbs}, {Tonry}, {White}, {Ba{\~n}ados}, {Bell},
  {Bender}, {Bernard}, {Boegner}, {Boffi}, {Botticella}, {Calamida},
  {Casertano}, {Chen}, {Chen}, {Cole}, {Deacon}, {Frenk}, {Fitzsimmons},
  {Gezari}, {Gibbs}, {Goessl}, {Goggia}, {Gourgue}, {Goldman}, {Grant},
  {Grebel}, {Hambly}, {Hasinger}, {Heavens}, {Heckman}, {Henderson}, {Henning},
  {Holman}, {Hopp}, {Ip}, {Isani}, {Jackson}, {Keyes}, {Koekemoer}, {Kotak},
  {Le}, {Liska}, {Long}, {Lucey}, {Liu}, {Martin}, {Masci}, {McLean}, {Mindel},
  {Misra}, {Morganson}, {Murphy}, {Obaika}, {Narayan}, {Nieto-Santisteban},
  {Norberg}, {Peacock}, {Pier}, {Postman}, {Primak}, {Rae}, {Rai}, {Riess},
  {Riffeser}, {Rix}, {R{\"o}ser}, {Russel}, {Rutz}, {Schilbach}, {Schultz},
  {Scolnic}, {Strolger}, {Szalay}, {Seitz}, {Small}, {Smith}, {Soderblom},
  {Taylor}, {Thomson}, {Taylor}, {Thakar}, {Thiel}, {Thilker}, {Unger},
  {Urata}, {Valenti}, {Wagner}, {Walder}, {Walter}, {Watters}, {Werner},
  {Wood-Vasey}, \& {Wyse}}]{chanbers2016panstarrs}
{Chambers}, K.~C., {Magnier}, E.~A., {Metcalfe}, N., {et~al.} 2016, arXiv
  e-prints, arXiv:1612.05560

\bibitem[{{Cushing} {et~al.}(2011{\natexlab{a}}){Cushing}, {Kirkpatrick},
  {Gelino}, {Griffith}, {Skrutskie}, {Mainzer}, {Marsh}, {Beichman},
  {Burgasser}, {Prato}, {Simcoe}, {Marley}, {Saumon}, {Freedman}, {Eisenhardt},
  \& {Wright}}]{cushing2011Y}
{Cushing}, M.~C., {Kirkpatrick}, J.~D., {Gelino}, C.~R., {et~al.}
  2011{\natexlab{a}}, \apj, 743, 50

\bibitem[{{Cushing} {et~al.}(2011{\natexlab{b}}){Cushing}, {Kirkpatrick},
  {Gelino}, {Griffith}, {Skrutskie}, {Mainzer}, {Marsh}, {Beichman},
  {Burgasser}, {Prato}, {Simcoe}, {Marley}, {Saumon}, {Freedman}, {Eisenhardt},
  \& {Wright}}]{cushing2011dY_WISE}
{Cushing}, M.~C., {Kirkpatrick}, J.~D., {Gelino}, C.~R., {et~al.}
  2011{\natexlab{b}}, \apj, 743, 50

\bibitem[{{Cushing} {et~al.}(2005){Cushing}, {Rayner}, \&
  {Vacca}}]{cushing2005IR_MLT}
{Cushing}, M.~C., {Rayner}, J.~T., \& {Vacca}, W.~D. 2005, \apj, 623, 1115

\bibitem[{{Cushing} {et~al.}(2021){Cushing}, {Schneider}, {Kirkpatrick},
  {Morley}, {Marley}, {Gelino}, {Mace}, {Wright}, {Eisenhardt}, {Skrutskie}, \&
  {Marsh}}]{cushing2021W1828}
{Cushing}, M.~C., {Schneider}, A.~C., {Kirkpatrick}, J.~D., {et~al.} 2021,
  \apj, 920, 20

\bibitem[{{Cutri} {et~al.}(2003){Cutri}, {Skrutskie}, {van Dyk}, {Beichman},
  {Carpenter}, {Chester}, {Cambresy}, {Evans}, {Fowler}, {Gizis}, {Howard},
  {Huchra}, {Jarrett}, {Kopan}, {Kirkpatrick}, {Light}, {Marsh}, {McCallon},
  {Schneider}, {Stiening}, {Sykes}, {Weinberg}, {Wheaton}, {Wheelock}, \&
  {Zacarias}}]{cutri2003twomasss_point_catalog}
{Cutri}, R.~M., {Skrutskie}, M.~F., {van Dyk}, S., {et~al.} 2003, VizieR Online
  Data Catalog, II/246

\bibitem[{{Cutri} {et~al.}(2012){Cutri}, {Wright}, {Conrow}, {Bauer},
  {Benford}, {Brandenburg}, {Dailey}, {Eisenhardt}, {Evans}, {Fajardo-Acosta},
  {Fowler}, {Gelino}, {Grillmair}, {Harbut}, {Hoffman}, {Jarrett},
  {Kirkpatrick}, {Leisawitz}, {Liu}, {Mainzer}, {Marsh}, {Masci}, {McCallon},
  {Padgett}, {Ressler}, {Royer}, {Skrutskie}, {Stanford}, {Wyatt}, {Tholen},
  {Tsai}, {Wachter}, {Wheelock}, {Yan}, {Alles}, {Beck}, {Grav}, {Masiero},
  {McCollum}, {McGehee}, {Papin}, \& {Wittman}}]{cutri2012wise}
{Cutri}, R.~M., {Wright}, E.~L., {Conrow}, T., {et~al.} 2012, {Explanatory
  Supplement to the WISE All-Sky Data Release Products}, Explanatory Supplement
  to the WISE All-Sky Data Release Products

\bibitem[{{Delorme} {et~al.}(2008){Delorme}, {Delfosse}, {Albert}, {Artigau},
  {Forveille}, {Reyl{\'e}}, {Allard}, {Homeier}, {Robin}, {Willott}, {Liu}, \&
  {Dupuy}}]{delorme2008TY_J0059}
{Delorme}, P., {Delfosse}, X., {Albert}, L., {et~al.} 2008, \aap, 482, 961

\bibitem[{{Dey} {et~al.}(2016){Dey}, {Rabinowitz}, {Karcher}, {Bebek},
  {Baltay}, {Sprayberry}, {Valdes}, {Stupak}, {Donaldson}, {Emmet}, {Hurteau},
  {Abareshi}, {Marshall}, {Lang}, {Fitzpatrick}, {Daly}, {Joyce}, {Schlegel},
  {Schweiker}, {Allen}, {Blum}, \& {Levi}}]{Dey2016kittpeak_mosaic3}
{Dey}, A., {Rabinowitz}, D., {Karcher}, A., {et~al.} 2016, in Society of
  Photo-Optical Instrumentation Engineers (SPIE) Conference Series, Vol. 9908,
  Ground-based and Airborne Instrumentation for Astronomy VI, ed. C.~J.
  {Evans}, L.~{Simard}, \& H.~{Takami}, 99082C

\bibitem[{{Dupuy} \& {Liu}(2017)}]{dupuy_liu2017dynamical_mass}
{Dupuy}, T.~J. \& {Liu}, M.~C. 2017, \apjs, 231, 15

\bibitem[{{Flaugher} {et~al.}(2015){Flaugher}, {Diehl}, {Honscheid}, {Abbott},
  {Alvarez}, {Angstadt}, {Annis}, {Antonik}, {Ballester}, {Beaufore},
  {Bernstein}, {Bernstein}, {Bigelow}, {Bonati}, {Boprie}, {Brooks},
  {Buckley-Geer}, {Campa}, {Cardiel-Sas}, {Castander}, {Castilla}, {Cease},
  {Cela-Ruiz}, {Chappa}, {Chi}, {Cooper}, {da Costa}, {Dede}, {Derylo},
  {DePoy}, {de Vicente}, {Doel}, {Drlica-Wagner}, {Eiting}, {Elliott}, {Emes},
  {Estrada}, {Fausti Neto}, {Finley}, {Flores}, {Frieman}, {Gerdes},
  {Gladders}, {Gregory}, {Gutierrez}, {Hao}, {Holland}, {Holm}, {Huffman},
  {Jackson}, {James}, {Jonas}, {Karcher}, {Karliner}, {Kent}, {Kessler},
  {Kozlovsky}, {Kron}, {Kubik}, {Kuehn}, {Kuhlmann}, {Kuk}, {Lahav}, {Lathrop},
  {Lee}, {Levi}, {Lewis}, {Li}, {Mandrichenko}, {Marshall}, {Martinez},
  {Merritt}, {Miquel}, {Mu{\~n}oz}, {Neilsen}, {Nichol}, {Nord}, {Ogando},
  {Olsen}, {Palaio}, {Patton}, {Peoples}, {Plazas}, {Rauch}, {Reil}, {Rheault},
  {Roe}, {Rogers}, {Roodman}, {Sanchez}, {Scarpine}, {Schindler}, {Schmidt},
  {Schmitt}, {Schubnell}, {Schultz}, {Schurter}, {Scott}, {Serrano}, {Shaw},
  {Smith}, {Soares-Santos}, {Stefanik}, {Stuermer}, {Suchyta}, {Sypniewski},
  {Tarle}, {Thaler}, {Tighe}, {Tran}, {Tucker}, {Walker}, {Wang}, {Watson},
  {Weaverdyck}, {Wester}, {Woods}, {Yanny}, \& {DES
  Collaboration}}]{flaugher2015DEScamara}
{Flaugher}, B., {Diehl}, H.~T., {Honscheid}, K., {et~al.} 2015, \aj, 150, 150

\bibitem[{{Geballe} {et~al.}(2002){Geballe}, {Knapp}, {Leggett}, {Fan},
  {Golimowski}, {Anderson}, {Brinkmann}, {Csabai}, {Gunn}, {Hawley},
  {Hennessy}, {Henry}, {Hill}, {Hindsley}, {Ivezi{\'c}}, {Lupton}, {McDaniel},
  {Munn}, {Narayanan}, {Peng}, {Pier}, {Rockosi}, {Schneider}, {Smith},
  {Strauss}, {Tsvetanov}, {Uomoto}, {York}, \&
  {Zheng}}]{geballe2002dT_classification}
{Geballe}, T.~R., {Knapp}, G.~R., {Leggett}, S.~K., {et~al.} 2002, \apj, 564,
  466

\bibitem[{{Gerasimov} {et~al.}(2020){Gerasimov}, {Homeier}, {Burgasser}, \&
  {Bedin}}]{gerasimov2020pheonix_metalpoor}
{Gerasimov}, R., {Homeier}, D., {Burgasser}, A., \& {Bedin}, L.~R. 2020,
  Research Notes of the American Astronomical Society, 4, 214

\bibitem[{{Greco} {et~al.}(2019){Greco}, {Schneider}, {Cushing}, {Kirkpatrick},
  \& {Burgasser}}]{greco2019neowise}
{Greco}, J.~J., {Schneider}, A.~C., {Cushing}, M.~C., {Kirkpatrick}, J.~D., \&
  {Burgasser}, A.~J. 2019, \aj, 158, 182

\bibitem[{{Hauschildt} \& {Baron}(1999)}]{hauschildt1999expand_atm_numerical}
{Hauschildt}, P.~H. \& {Baron}, E. 1999, Journal of Computational and Applied
  Mathematics, 109, 41

\bibitem[{{Kirkpatrick} {et~al.}(2011){Kirkpatrick}, {Cushing}, {Gelino},
  {Griffith}, {Skrutskie}, {Marsh}, {Wright}, {Mainzer}, {Eisenhardt},
  {McLean}, {Thompson}, {Bauer}, {Benford}, {Bridge}, {Lake}, {Petty},
  {Stanford}, {Tsai}, {Bailey}, {Beichman}, {Bloom}, {Bochanski}, {Burgasser},
  {Capak}, {Cruz}, {Hinz}, {Kartaltepe}, {Knox}, {Manohar}, {Masters},
  {Morales-Calder{\'o}n}, {Prato}, {Rodigas}, {Salvato}, {Schurr}, {Scoville},
  {Simcoe}, {Stapelfeldt}, {Stern}, {Stock}, \&
  {Vacca}}]{kirkpatrick2011hundredBD_WISE}
{Kirkpatrick}, J.~D., {Cushing}, M.~C., {Gelino}, C.~R., {et~al.} 2011, \apjs,
  197, 19

\bibitem[{{Kirkpatrick} {et~al.}(2019){Kirkpatrick}, {Martin}, {Smart},
  {Cayago}, {Beichman}, {Marocco}, {Gelino}, {Faherty}, {Cushing}, {Schneider},
  {Mace}, {Tinney}, {Wright}, {Lowrance}, {Ingalls}, {Vrba}, {Munn}, {Dahm}, \&
  {McLean}}]{kirkpatrick2019parallax184TY}
{Kirkpatrick}, J.~D., {Martin}, E.~C., {Smart}, R.~L., {et~al.} 2019, \apjs,
  240, 19

\bibitem[{{Kirkpatrick} {et~al.}(1999){Kirkpatrick}, {Reid}, {Liebert},
  {Cutri}, {Nelson}, {Beichman}, {Dahn}, {Monet}, {Gizis}, \&
  {Skrutskie}}]{kirkpatrick1999L}
{Kirkpatrick}, J.~D., {Reid}, I.~N., {Liebert}, J., {et~al.} 1999, \apj, 519,
  802

\bibitem[{Kumar(1963)}]{kumar1963structureBD}
Kumar, S.~S. 1963, The Astrophysical Journal, 137, 1121

\bibitem[{{Lang} {et~al.}(2010){Lang}, {Hogg}, {Mierle}, {Blanton}, \&
  {Roweis}}]{dustin2010astrometry.net}
{Lang}, D., {Hogg}, D.~W., {Mierle}, K., {Blanton}, M., \& {Roweis}, S. 2010,
  \aj, 139, 1782

\bibitem[{{Lawrence} {et~al.}(2007{\natexlab{a}}){Lawrence}, {Warren},
  {Almaini}, {Edge}, {Hambly}, {Jameson}, {Lucas}, {Casali}, {Adamson}, {Dye},
  {Emerson}, {Foucaud}, {Hewett}, {Hirst}, {Hodgkin}, {Irwin}, {Lodieu},
  {McMahon}, {Simpson}, {Smail}, {Mortlock}, \& {Folger}}]{lawrence2007ulas}
{Lawrence}, A., {Warren}, S.~J., {Almaini}, O., {et~al.} 2007{\natexlab{a}},
  \mnras, 379, 1599

\bibitem[{{Lawrence} {et~al.}(2007{\natexlab{b}}){Lawrence}, {Warren},
  {Almaini}, {Edge}, {Hambly}, {Jameson}, {Lucas}, {Casali}, {Adamson}, {Dye},
  {Emerson}, {Foucaud}, {Hewett}, {Hirst}, {Hodgkin}, {Irwin}, {Lodieu},
  {McMahon}, {Simpson}, {Smail}, {Mortlock}, \& {Folger}}]{lawrence2007ukidss}
{Lawrence}, A., {Warren}, S.~J., {Almaini}, O., {et~al.} 2007{\natexlab{b}},
  \mnras, 379, 1599

\bibitem[{{Lodieu} {et~al.}(2019){Lodieu}, {Allard}, {Rodrigo}, {Pavlenko},
  {Burgasser}, {Lyubchik}, {Kaminsky}, \& {Homeier}}]{lodieu2019sdM}
{Lodieu}, N., {Allard}, F., {Rodrigo}, C., {et~al.} 2019, \aap, 628, A61

\bibitem[{{Lodieu} {et~al.}(2013){Lodieu}, {B{\'e}jar}, \&
  {Rebolo}}]{lodieu2013Yoptical}
{Lodieu}, N., {B{\'e}jar}, V.~J.~S., \& {Rebolo}, R. 2013, \aap, 550, L2

\bibitem[{{Lodieu} {et~al.}(2022){Lodieu}, {Zapatero Osorio}, {Mart{\'\i}n},
  {Rebolo L{\'o}pez}, \& {Gauza}}]{lodieu2022W1810}
{Lodieu}, N., {Zapatero Osorio}, M.~R., {Mart{\'\i}n}, E.~L., {Rebolo
  L{\'o}pez}, R., \& {Gauza}, B. 2022, \aap, 663, A84

\bibitem[{{Lodieu} {et~al.}(2015){Lodieu}, {Zapatero Osorio}, {Rebolo},
  {B{\'e}jar}, {Pavlenko}, \& {P{\'e}rez-Garrido}}]{lodieu2015luhman16xshooter}
{Lodieu}, N., {Zapatero Osorio}, M.~R., {Rebolo}, R., {et~al.} 2015, \aap, 581,
  A73

\bibitem[{{Marocco} {et~al.}(2021){Marocco}, {Eisenhardt}, {Fowler},
  {Kirkpatrick}, {Meisner}, {Schlafly}, {Stanford}, {Garcia}, {Caselden},
  {Cushing}, {Cutri}, {Faherty}, {Gelino}, {Gonzalez}, {Jarrett}, {Koontz},
  {Mainzer}, {Marchese}, {Mobasher}, {Schlegel}, {Stern}, {Teplitz}, \&
  {Wright}}]{marocco2021catwise}
{Marocco}, F., {Eisenhardt}, P. R.~M., {Fowler}, J.~W., {et~al.} 2021, \apjs,
  253, 8

\bibitem[{{Mart{\'\i}n} {et~al.}(1999){Mart{\'\i}n}, {Delfosse}, {Basri},
  {Goldman}, {Forveille}, \& {Zapatero Osorio}}]{martin1999Lclassification}
{Mart{\'\i}n}, E.~L., {Delfosse}, X., {Basri}, G., {et~al.} 1999, \aj, 118,
  2466

\bibitem[{{Mart{\'\i}n} {et~al.}(2021){Mart{\'\i}n}, {Zhang}, {Esparza},
  {Gracia}, {Rasilla}, {Masseron}, \& {Burgasser}}]{martin2021ch4nh3}
{Mart{\'\i}n}, E.~L., {Zhang}, J.~Y., {Esparza}, P., {et~al.} 2021, \aap, 655,
  L3

\bibitem[{{Martín}(2023)}]{martin2023Yoptical}
{Martín}, E.~L. 2023, in prep.

\bibitem[{{Meisner} {et~al.}(2020){Meisner}, {Faherty}, {Kirkpatrick},
  {Schneider}, {Caselden}, {Gagn{\'e}}, {Kuchner}, {Burgasser}, {Casewell},
  {Debes}, {Artigau}, {Bardalez Gagliuffi}, {Logsdon}, {Kiman}, {Allers},
  {Hsu}, {Wisniewski}, {Allen}, {Beaulieu}, {Colin}, {Durantini Luca},
  {Goodman}, {Gramaize}, {Hamlet}, {Hinckley}, {Kiwy}, {Martin}, {Pendrill},
  {Rothermich}, {Sainio}, {Sch{\"u}mann}, {Andersen}, {Tanner}, {Thakur},
  {Th{\'e}venot}, {Walla}, {W{\k{e}}dracki}, {Aganze}, {Gerasimov}, {Theissen},
  \& {Backyard Worlds: Planet 9 Collaboration}}]{meisner2020extremecoldBD}
{Meisner}, A.~M., {Faherty}, J.~K., {Kirkpatrick}, J.~D., {et~al.} 2020, \apj,
  899, 123

\bibitem[{{Meisner} {et~al.}(2023){Meisner}, {Leggett}, {Logsdon}, {Schneider},
  {Tremblin}, \& {Phillips}}]{meisner2023coldoldBD}
{Meisner}, A.~M., {Leggett}, S.~K., {Logsdon}, S.~E., {et~al.} 2023, arXiv
  e-prints, arXiv:2301.09817

\bibitem[{{Meisner} {et~al.}(2021){Meisner}, {Schneider}, {Burgasser},
  {Marocco}, {Line}, {Faherty}, {Kirkpatrick}, {Caselden}, {Kuchner}, {Gelino},
  {Gagn{\'e}}, {Theissen}, {Gerasimov}, {Aganze}, {Hsu}, {Wisniewski},
  {Casewell}, {Bardalez Gagliuffi}, {Logsdon}, {Eisenhardt}, {Allers}, {Debes},
  {Allen}, {Stevnbak Andersen}, {Goodman}, {Gramaize}, {Martin}, {Sainio},
  {Cushing}, \& {Backyard Worlds: Planet 9 Collaboration}}]{meisner2021esdT}
{Meisner}, A.~M., {Schneider}, A.~C., {Burgasser}, A.~J., {et~al.} 2021, \apj,
  915, 120

\bibitem[{{Morganson} {et~al.}(2018){Morganson}, {Gruendl}, {Menanteau},
  {Carrasco Kind}, {Chen}, {Daues}, {Drlica-Wagner}, {Friedel}, {Gower},
  {Johnson}, {Johnson}, {Kessler}, {Paz-Chinch{\'o}n}, {Petravick}, {Pond},
  {Yanny}, {Allam}, {Armstrong}, {Barkhouse}, {Bechtol}, {Benoit-L{\'e}vy},
  {Bernstein}, {Bertin}, {Buckley-Geer}, {Covarrubias}, {Desai}, {Diehl},
  {Goldstein}, {Gruen}, {Li}, {Lin}, {Marriner}, {Mohr}, {Neilsen}, {Ngeow},
  {Paech}, {Rykoff}, {Sako}, {Sevilla-Noarbe}, {Sheldon}, {Sobreira}, {Tucker},
  {Wester}, \& {DES Collaboration}}]{morganson2018DESpipeline}
{Morganson}, E., {Gruendl}, R.~A., {Menanteau}, F., {et~al.} 2018, \pasp, 130,
  074501

\bibitem[{{Morley} {et~al.}(2014){Morley}, {Marley}, {Fortney}, {Lupu},
  {Saumon}, {Greene}, \& {Lodders}}]{morley2014Ywaterclouds}
{Morley}, C.~V., {Marley}, M.~S., {Fortney}, J.~J., {et~al.} 2014, \apj, 787,
  78

\bibitem[{{Murray} {et~al.}(2011){Murray}, {Burningham}, {Jones}, {Pinfield},
  {Lucas}, {Leggett}, {Tinney}, {Day-Jones}, {Weights}, {Lodieu}, {P{\'e}rez
  Prieto}, {Nickson}, {Zhang}, {Clarke}, {Jenkins}, \&
  {Tamura}}]{murray_burningham2011blueT}
{Murray}, D.~N., {Burningham}, B., {Jones}, H.~R.~A., {et~al.} 2011, \mnras,
  414, 575

\bibitem[{Nakajima {et~al.}(1995)Nakajima, Oppenheimer, Kulkarni, Golimowski,
  Matthews, \& Durrance}]{nakajima1995discovery}
Nakajima, T., Oppenheimer, B., Kulkarni, S., {et~al.} 1995, nature, 378, 463

\bibitem[{{Nidever} {et~al.}(2018){Nidever}, {Dey}, {Olsen}, {Ridgway},
  {Nikutta}, {Juneau}, {Fitzpatrick}, {Scott}, \&
  {Valdes}}]{nidever2018_survey_w1553zmag}
{Nidever}, D.~L., {Dey}, A., {Olsen}, K., {et~al.} 2018, \aj, 156, 131

\bibitem[{{Oke}(1974)}]{oke1974ABsystem}
{Oke}, J.~B. 1974, \apjs, 27, 21

\bibitem[{{Pinfield} {et~al.}(2014){Pinfield}, {Gomes}, {Day-Jones}, {Leggett},
  {Gromadzki}, {Burningham}, {Ruiz}, {Kurtev}, {Cattermole}, {Cardoso},
  {Lodieu}, {Faherty}, {Littlefair}, {Smart}, {Irwin}, {Clarke}, {Smith},
  {Lucas}, {G{\'a}lvez-Ortiz}, {Jenkins}, {Jones}, {Rebolo}, {B{\'e}jar}, \&
  {Gauza}}]{pinfield2014subdwarf}
{Pinfield}, D.~J., {Gomes}, J., {Day-Jones}, A.~C., {et~al.} 2014, \mnras, 437,
  1009

\bibitem[{{Prochaska} {et~al.}(2020){Prochaska}, {Hennawi}, {Cooke},
  {Westfall}, {Wang}, {EmAstro}, {Tiffanyhsyu}, {Wasserman}, {Villaume},
  {Marijana777}, {Schindler}, {Young}, {Simha}, {Wilde}, {Tejos}, {Isbell},
  {Fl{\"o}rs}, {Sandford}, {Vasovi{\'c}}, {Betts}, \& {Holden}}]{pypeit:zenodo}
{Prochaska}, J.~X., {Hennawi}, J., {Cooke}, R., {et~al.} 2020, {pypeit/PypeIt:
  Release 1.0.0}

\bibitem[{Prochaska {et~al.}(2020)Prochaska, Hennawi, Westfall, Cooke, Wang,
  Hsyu, Davies, Farina, \& Pelliccia}]{pypeit:joss_pub}
Prochaska, J.~X., Hennawi, J.~F., Westfall, K.~B., {et~al.} 2020, Journal of
  Open Source Software, 5, 2308

\bibitem[{Rebolo {et~al.}(1992)Rebolo, Martin, \&
  Magazzu}]{rebolo1992spectroscopy_lithum}
Rebolo, R., Martin, E.~L., \& Magazzu, A. 1992, The Astrophysical Journal, 389,
  L83

\bibitem[{Rebolo {et~al.}(1995)Rebolo, Osorio, \&
  Mart{\'\i}n}]{rebolo1995discovery}
Rebolo, R., Osorio, M., \& Mart{\'\i}n, E. 1995, Nature, 377, 129

\bibitem[{{Schneider} {et~al.}(2020){Schneider}, {Burgasser}, {Gerasimov},
  {Marocco}, {Gagn{\'e}}, {Goodman}, {Beaulieu}, {Pendrill}, {Rothermich},
  {Sainio}, {Kuchner}, {Caselden}, {Meisner}, {Faherty}, {Mamajek}, {Hsu},
  {Greco}, {Cushing}, {Kirkpatrick}, {Bardalez-Gagliuffi}, {Logsdon}, {Allers},
  {Debes}, \& {Backyard Worlds: Planet 9
  Collaboration}}]{Schneider2020W0414_W1810}
{Schneider}, A.~C., {Burgasser}, A.~J., {Gerasimov}, R., {et~al.} 2020, \apj,
  898, 77

\bibitem[{{Schneider} {et~al.}(2015){Schneider}, {Cushing}, {Kirkpatrick},
  {Gelino}, {Mace}, {Wright}, {Eisenhardt}, {Skrutskie}, {Griffith}, \&
  {Marsh}}]{Schneider2015HubbleWISE_BD}
{Schneider}, A.~C., {Cushing}, M.~C., {Kirkpatrick}, J.~D., {et~al.} 2015,
  \apj, 804, 92

\bibitem[{{Skrutskie} {et~al.}(2006){Skrutskie}, {Cutri}, {Stiening},
  {Weinberg}, {Schneider}, {Carpenter}, {Beichman}, {Capps}, {Chester},
  {Elias}, {Huchra}, {Liebert}, {Lonsdale}, {Monet}, {Price}, {Seitzer},
  {Jarrett}, {Kirkpatrick}, {Gizis}, {Howard}, {Evans}, {Fowler}, {Fullmer},
  {Hurt}, {Light}, {Kopan}, {Marsh}, {McCallon}, {Tam}, {Van Dyk}, \&
  {Wheelock}}]{skrutskie2006_2MASS}
{Skrutskie}, M.~F., {Cutri}, R.~M., {Stiening}, R., {et~al.} 2006, \aj, 131,
  1163

\bibitem[{{Tody}(1986)}]{tody1986iraf}
{Tody}, D. 1986, in Society of Photo-Optical Instrumentation Engineers (SPIE)
  Conference Series, Vol. 627, Instrumentation in astronomy VI, ed. D.~L.
  {Crawford}, 733

\bibitem[{{Wright} {et~al.}(2010){Wright}, {Eisenhardt}, {Mainzer}, {Ressler},
  {Cutri}, {Jarrett}, {Kirkpatrick}, {Padgett}, {McMillan}, {Skrutskie},
  {Stanford}, {Cohen}, {Walker}, {Mather}, {Leisawitz}, {Gautier}, {McLean},
  {Benford}, {Lonsdale}, {Blain}, {Mendez}, {Irace}, {Duval}, {Liu}, {Royer},
  {Heinrichsen}, {Howard}, {Shannon}, {Kendall}, {Walsh}, {Larsen}, {Cardon},
  {Schick}, {Schwalm}, {Abid}, {Fabinsky}, {Naes}, \& {Tsai}}]{wright2010WISE}
{Wright}, E.~L., {Eisenhardt}, P. R.~M., {Mainzer}, A.~K., {et~al.} 2010, \aj,
  140, 1868

\bibitem[{{York} {et~al.}(2000){York}, {Adelman}, {Anderson}, {Anderson},
  {Annis}, {Bahcall}, {Bakken}, {Barkhouser}, {Bastian}, {Berman}, {Boroski},
  {Bracker}, {Briegel}, {Briggs}, {Brinkmann}, {Brunner}, {Burles}, {Carey},
  {Carr}, {Castander}, {Chen}, {Colestock}, {Connolly}, {Crocker}, {Csabai},
  {Czarapata}, {Davis}, {Doi}, {Dombeck}, {Eisenstein}, {Ellman}, {Elms},
  {Evans}, {Fan}, {Federwitz}, {Fiscelli}, {Friedman}, {Frieman}, {Fukugita},
  {Gillespie}, {Gunn}, {Gurbani}, {de Haas}, {Haldeman}, {Harris}, {Hayes},
  {Heckman}, {Hennessy}, {Hindsley}, {Holm}, {Holmgren}, {Huang}, {Hull},
  {Husby}, {Ichikawa}, {Ichikawa}, {Ivezi{\'c}}, {Kent}, {Kim}, {Kinney},
  {Klaene}, {Kleinman}, {Kleinman}, {Knapp}, {Korienek}, {Kron}, {Kunszt},
  {Lamb}, {Lee}, {Leger}, {Limmongkol}, {Lindenmeyer}, {Long}, {Loomis},
  {Loveday}, {Lucinio}, {Lupton}, {MacKinnon}, {Mannery}, {Mantsch}, {Margon},
  {McGehee}, {McKay}, {Meiksin}, {Merelli}, {Monet}, {Munn}, {Narayanan},
  {Nash}, {Neilsen}, {Neswold}, {Newberg}, {Nichol}, {Nicinski}, {Nonino},
  {Okada}, {Okamura}, {Ostriker}, {Owen}, {Pauls}, {Peoples}, {Peterson},
  {Petravick}, {Pier}, {Pope}, {Pordes}, {Prosapio}, {Rechenmacher}, {Quinn},
  {Richards}, {Richmond}, {Rivetta}, {Rockosi}, {Ruthmansdorfer}, {Sandford},
  {Schlegel}, {Schneider}, {Sekiguchi}, {Sergey}, {Shimasaku}, {Siegmund},
  {Smee}, {Smith}, {Snedden}, {Stone}, {Stoughton}, {Strauss}, {Stubbs},
  {SubbaRao}, {Szalay}, {Szapudi}, {Szokoly}, {Thakar}, {Tremonti}, {Tucker},
  {Uomoto}, {Vanden Berk}, {Vogeley}, {Waddell}, {Wang}, {Watanabe},
  {Weinberg}, {Yanny}, {Yasuda}, \& {SDSS Collaboration}}]{york2000SDSS}
{York}, D.~G., {Adelman}, J., {Anderson}, John~E., J., {et~al.} 2000, \aj, 120,
  1579

\bibitem[{{Zapatero Osorio} {et~al.}(2018){Zapatero Osorio}, {B{\'e}jar},
  {Lodieu}, \& {Manjavacas}}]{zapatero2018Pleiades_leastmassive}
{Zapatero Osorio}, M.~R., {B{\'e}jar}, V.~J.~S., {Lodieu}, N., \& {Manjavacas},
  E. 2018, \mnras, 475, 139

\bibitem[{{Zhang} {et~al.}(2019){Zhang}, {Luo}, {Comte}, {Gizis}, {Wang}, {Li},
  {Qin}, {Kong}, {Bai}, \& {Yi}}]{zhangshuo2019sdM_identification_sample}
{Zhang}, S., {Luo}, A.~L., {Comte}, G., {et~al.} 2019, \apjs, 240, 31

\bibitem[{{Zhang}(2019)}]{zhangzenghua2019MLbinary}
{Zhang}, Z. 2019, \mnras, 489, 1423

\bibitem[{{Zhang} {et~al.}(2018){Zhang}, {Galvez-Ortiz}, {Pinfield},
  {Burgasser}, {Lodieu}, {Jones}, {Mart{\'\i}n}, {Burningham}, {Homeier},
  {Allard}, {Zapatero Osorio}, {Smith}, {Smart}, {L{\'o}pez Mart{\'\i}},
  {Marocco}, \& {Rebolo}}]{zhangzenghua2018subL}
{Zhang}, Z.~H., {Galvez-Ortiz}, M.~C., {Pinfield}, D.~J., {et~al.} 2018,
  \mnras, 480, 5447

\bibitem[{{Zhang} {et~al.}(2017){Zhang}, {Pinfield}, {G{\'a}lvez-Ortiz},
  {Burningham}, {Lodieu}, {Marocco}, {Burgasser}, {Day-Jones}, {Allard},
  {Jones}, {Homeier}, {Gomes}, \& {Smart}}]{zhang2017six_sdL_classification}
{Zhang}, Z.~H., {Pinfield}, D.~J., {G{\'a}lvez-Ortiz}, M.~C., {et~al.} 2017,
  \mnras, 464, 3040

\end{thebibliography}

\onecolumn

\begin{appendix}
\section{$z$-band images of the targets}

\begin{figure}[htbp]
     \centering
     \begin{subfigure}[b]{0.226\textwidth}
         \centering
         \includegraphics[width=\textwidth]{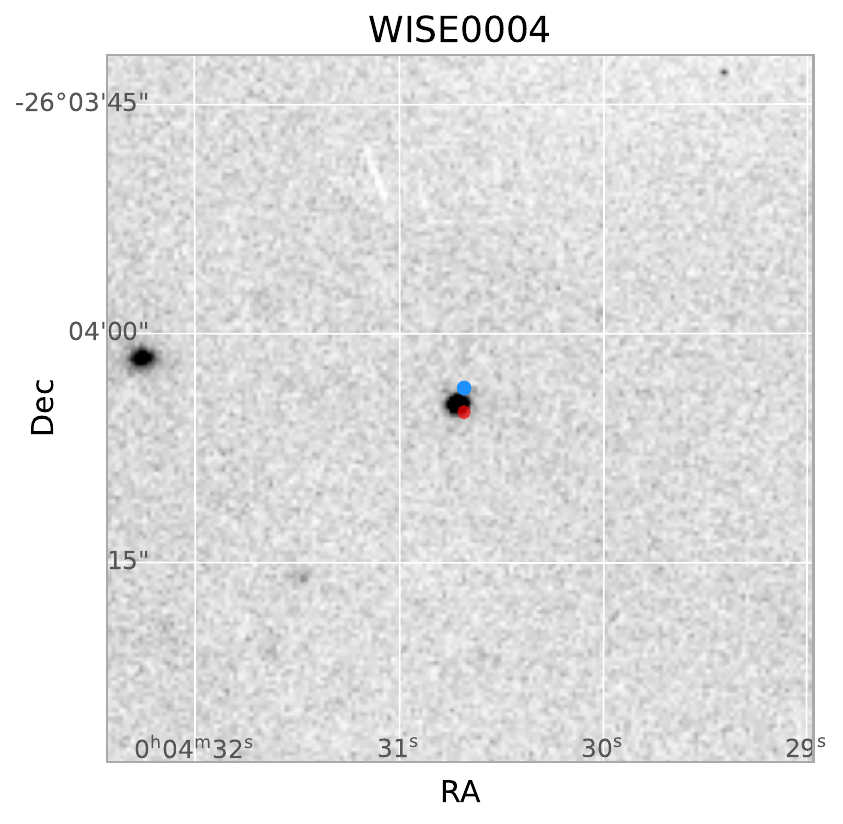}
         \label{W0004_illu}
     \end{subfigure}
     \begin{subfigure}[b]{0.216\textwidth}
         \centering
         \includegraphics[width=\textwidth]{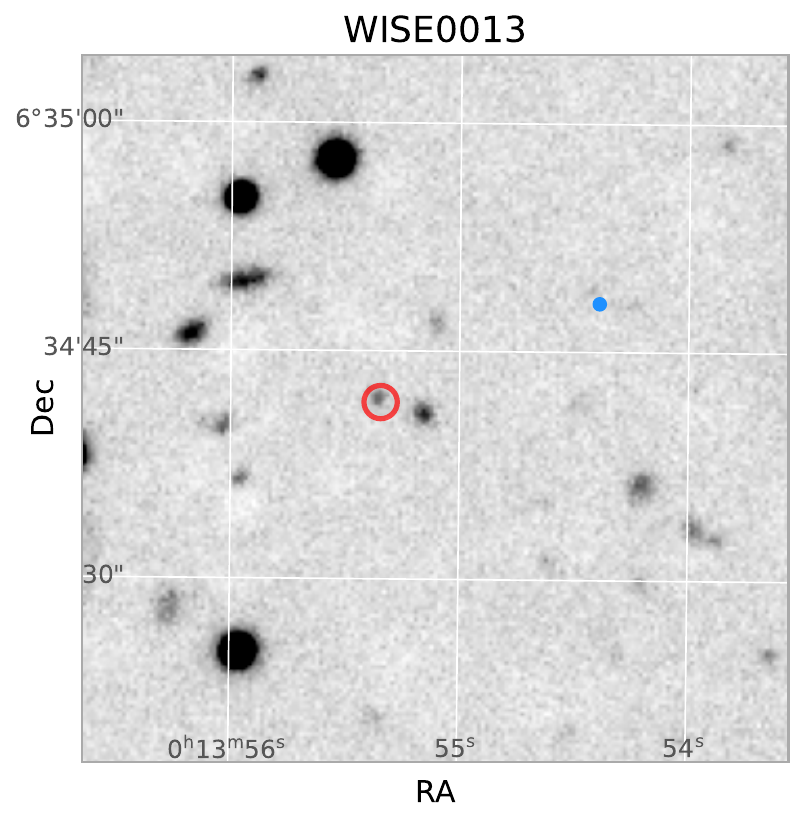}
         \label{W0013_illu}
     \end{subfigure}
     \begin{subfigure}[b]{0.225\textwidth}
         \centering
         \includegraphics[width=\textwidth]{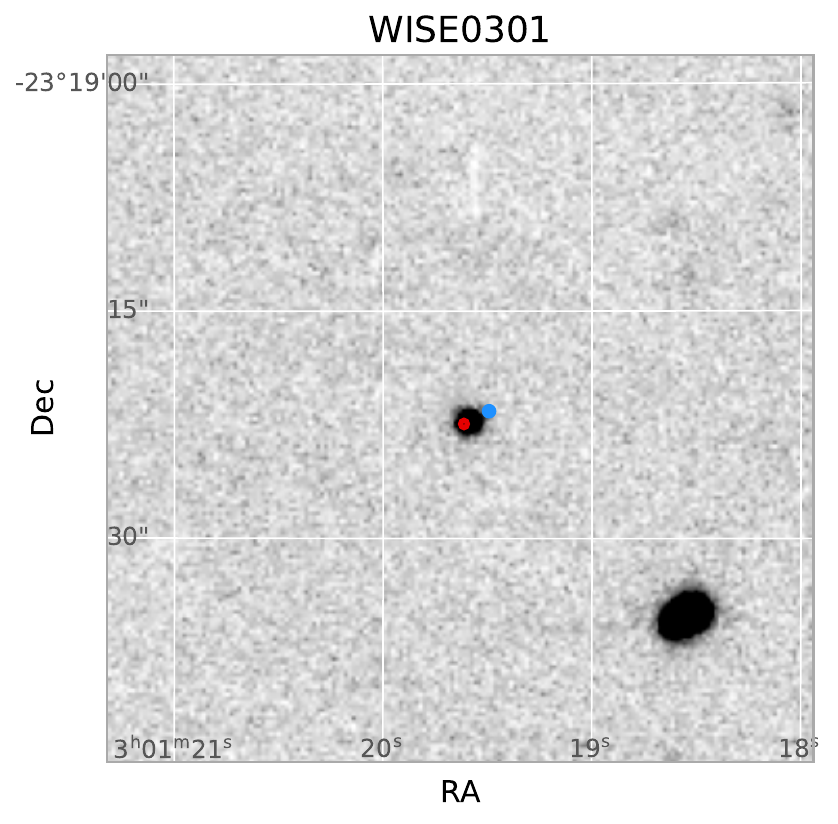}
         \label{W0301_illu}
     \end{subfigure}

     \begin{subfigure}[b]{0.22\textwidth}
         \centering
         \includegraphics[width=\textwidth]{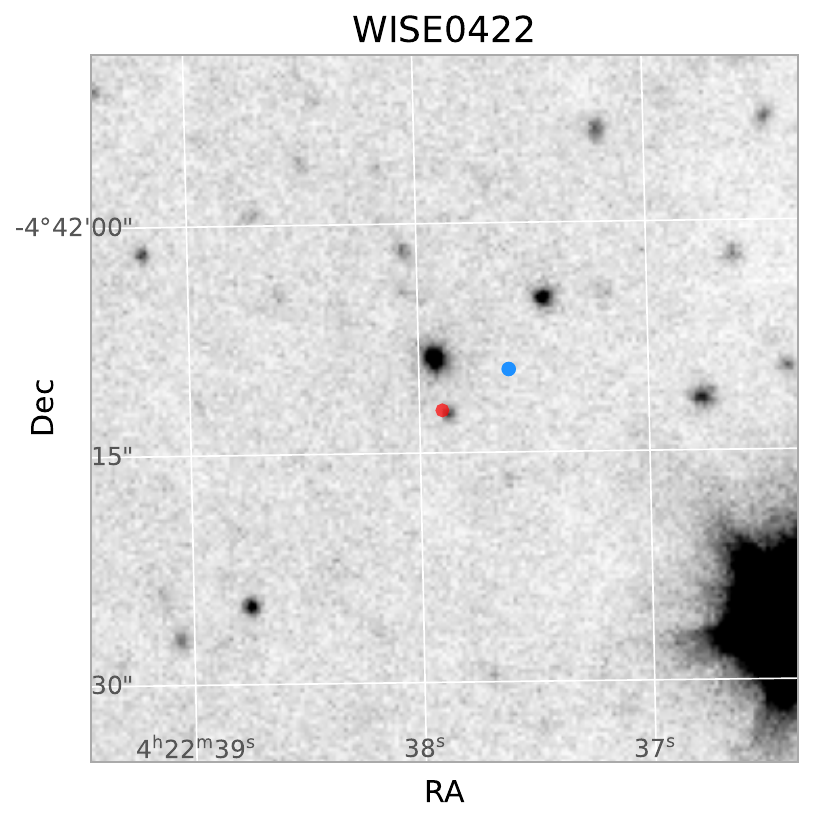}
         \label{W0422_illu}
     \end{subfigure}
     \begin{subfigure}[b]{0.218\textwidth}
         \centering
         \includegraphics[width=\textwidth]{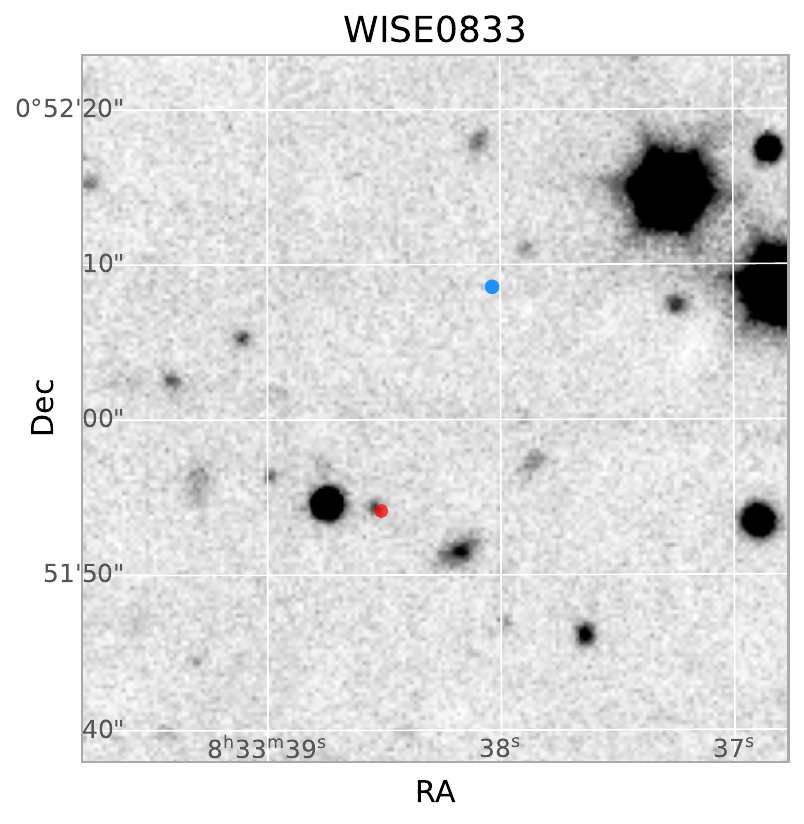}
         \label{W0833_illu}
     \end{subfigure}
        \begin{subfigure}[b]{0.218\textwidth}
         \centering
         \includegraphics[width=\textwidth]{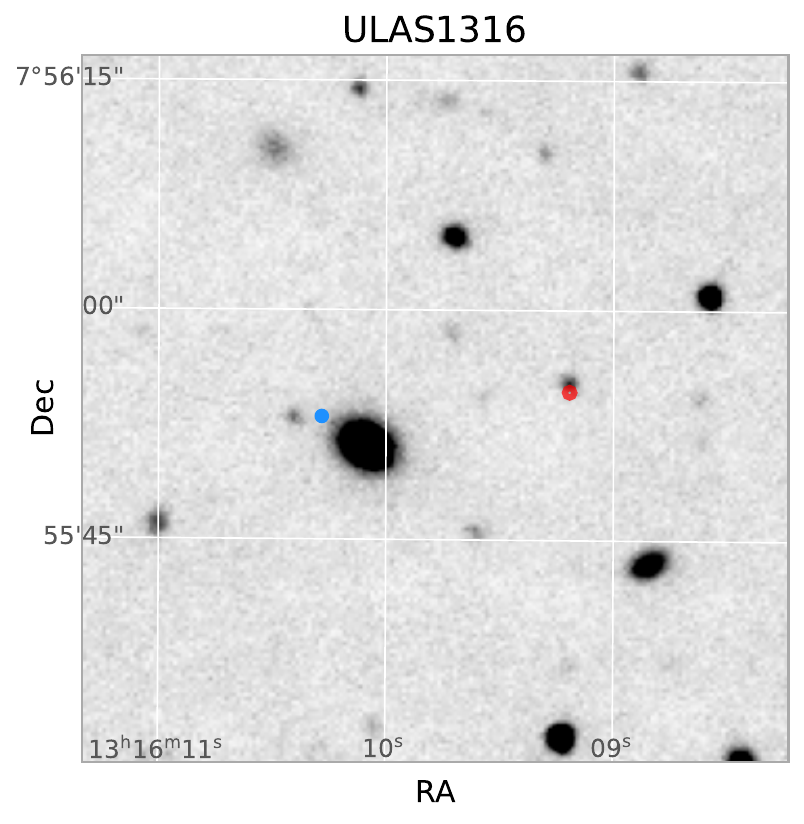}
         \label{ULAS1316_illu}
     \end{subfigure}
     \begin{subfigure}[b]{0.223\textwidth}
         \centering
         \includegraphics[width=\textwidth]{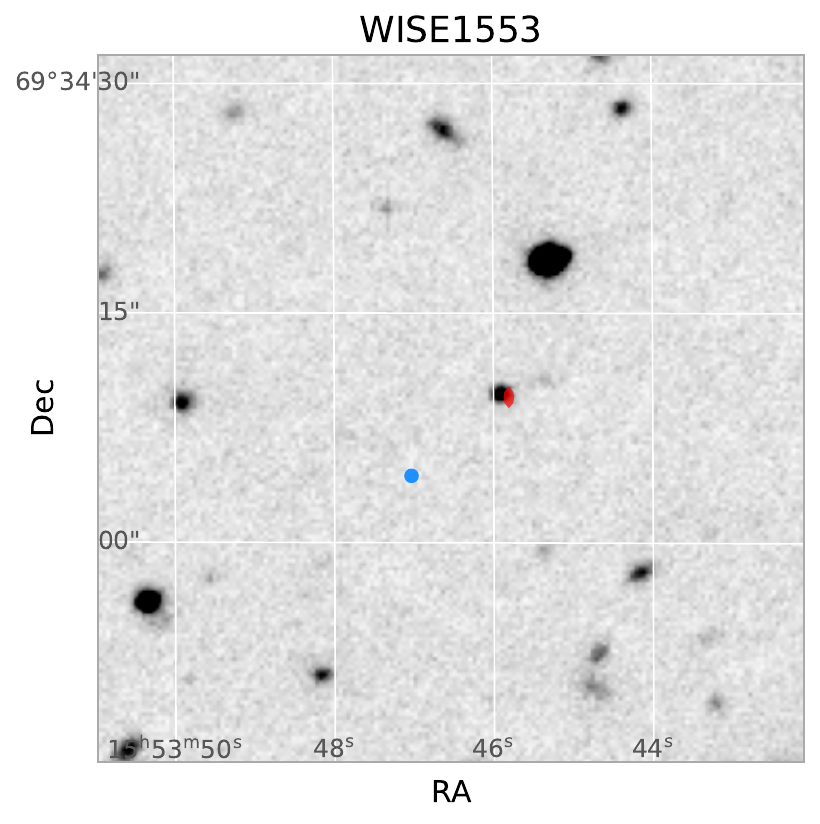}
         \label{W1553_illu}
     \end{subfigure}
     
         \begin{subfigure}[b]{0.218\textwidth}
         \centering
         \includegraphics[width=\textwidth]{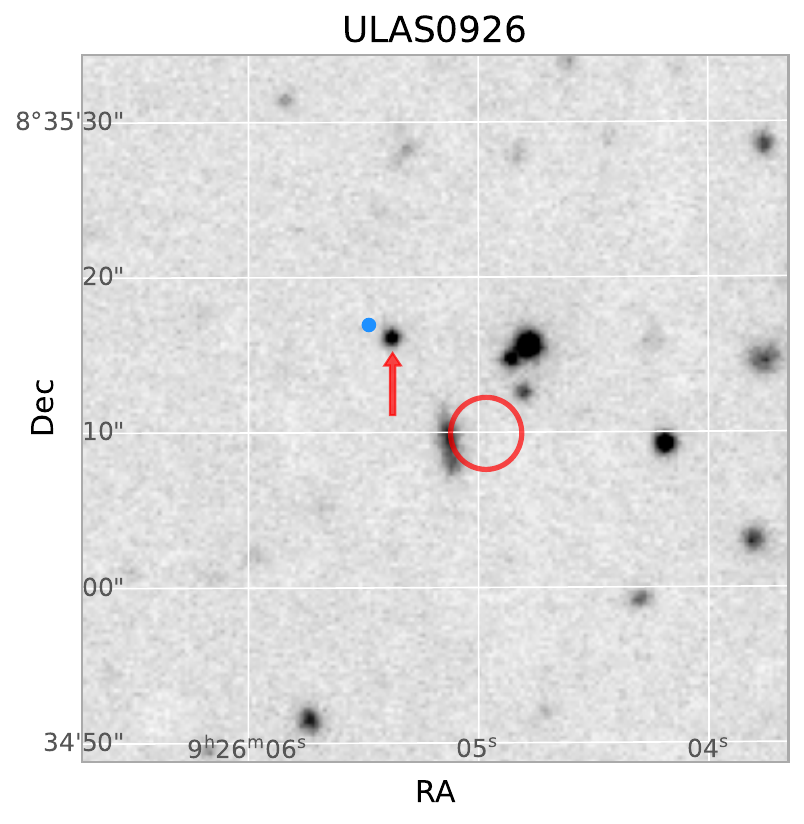}
         \label{ULAS0926_illu}
     \end{subfigure}
     \begin{subfigure}[b]{0.223\textwidth}
         \centering
         \includegraphics[width=\textwidth]{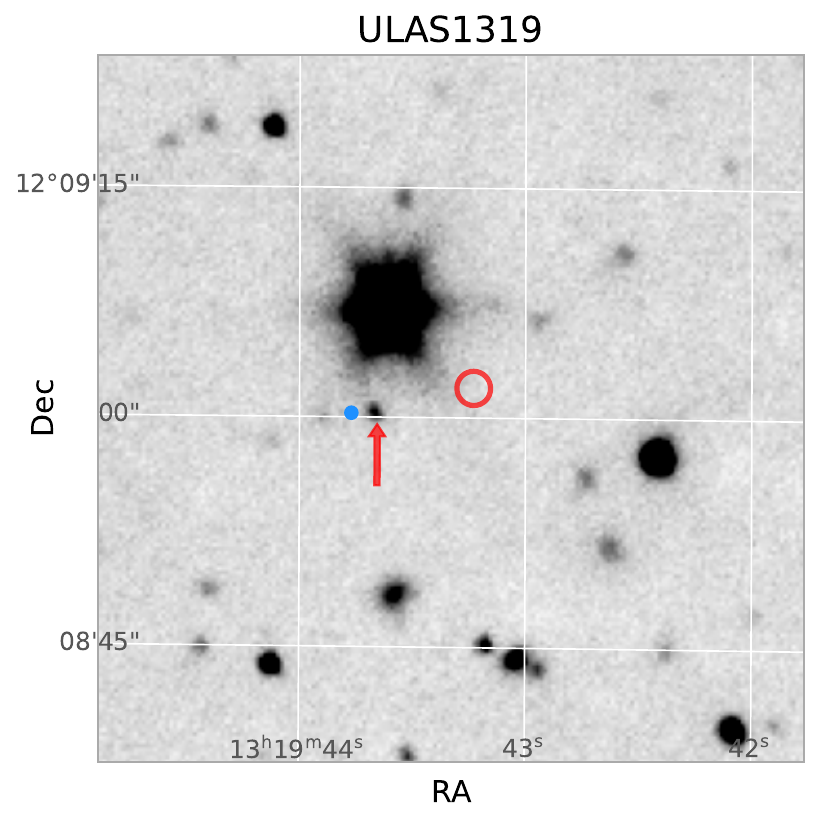}
         \label{ULAS1319_illu}
     \end{subfigure}
          \begin{subfigure}[b]{0.225\textwidth}
         \centering
         \includegraphics[width=\textwidth]{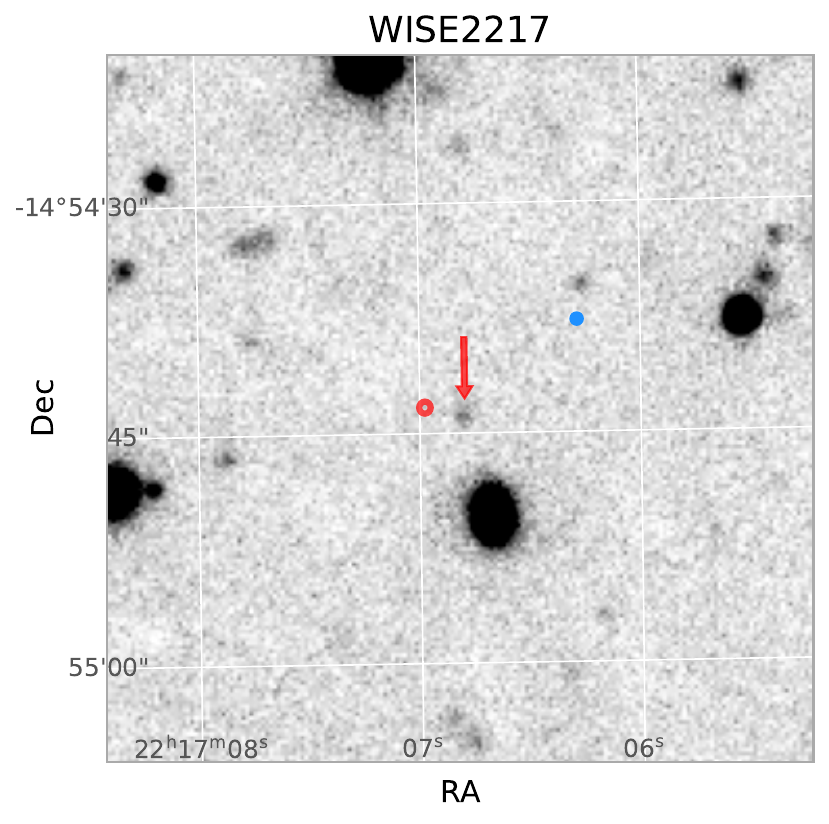}
         \label{W2217_illu}
     \end{subfigure}

        \caption{$45^\prime \times 45^\prime$ GTC OSIRIS 
$z$-band images of 
the fields of the all targets with the conventional direction 
(up is north and left is east). The blue dots are the object 
positions at the previous epochs published by other authors, 
(Table~1) and the red ellipses are the projected positions 
where those targets are supposed to be on the dates of the observations 
in Table~\ref{Obs} according to their proper motions from the 
literature in Table~\ref{tg}. The two semi-axes of the ellipses 
show the errors of the proper motions in RA and Dec. WISE0004, 
WISE0013, WISE0301, WISE0422, WISE0833, ULAS1316 and WISE1553 
were unambiguously detected within the error ellipses at their 
projected positions. However, for ULAS0926 and ULAS1319 we 
found the proper motions given in the literature are overestimated; 
and for WISE2217 the proper motion is deviated. The red arrows
 indicate the positions at which the three sources should be. We 
recalculated their proper motions according to these positions 
and we listed the results in Table~\ref{astrometry}.  }
        \label{fig:three graphs}
\end{figure}

\begin{figure}[htbp]
    \centering
         \includegraphics[width=0.228\textwidth]{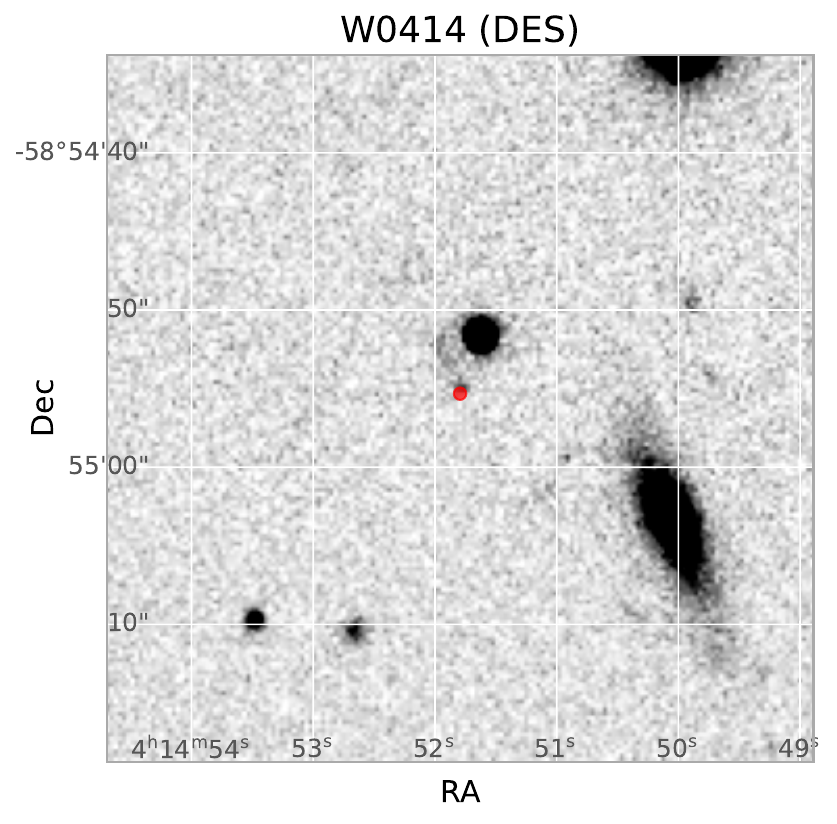}
         
    \caption{45' $\times$ 45' DES $z$-band field of WISE0414 
with the conventional direction (up is north and left is east). 
The red dot is the position WISE0414 in VHS $J$-band, whose 
observation epoch is only 27 days before DES observation epoch.}
    \label{W0414_illu}
\end{figure}

\end{appendix}

\end{document}